\documentclass[useAMS,usedcolumn,usenatbib,usegraphicx]{mn2e}
\usepackage{times}

\usepackage{aas_macros}
\usepackage{textcmds}
\usepackage{multirow}
\usepackage{xcolor}
\usepackage{hyperref}
\usepackage{ctable}
\usepackage{amsmath,amssymb}
\usepackage{diagbox, lscape, chemfig, subcaption}
\newcommand\T{\rule{0pt}{2.6ex}}       
\newcommand\B{\rule[-1.2ex]{0pt}{0pt}} 

\title[Cometary ices in forming disc midplanes]{Cometary ices in forming protoplanetary disc midplanes}

\author[Maria N. Drozdovskaya et al.]{Maria~N.~Drozdovskaya$^{1}$\thanks{E-mail: drozdovskaya@strw.leidenuniv.nl}, Catherine~Walsh$^{1}$, Ewine~F.~van~Dishoeck$^{1,2}$, Kenji Furuya$^{1}$,\newauthor
Ulysse Marboeuf$^{3,4}$, Amaury Thiabaud$^{3,4}$, Daniel~Harsono$^{5}$ and Ruud~Visser$^{6}$\\
$^{1}$~Leiden Observatory, Leiden University, P.O. Box 9513, 2300 RA, Leiden, The Netherlands\\
$^{2}$~Max-Planck-Institut f\"{u}r Extraterrestrische Physik, Giessenbachstrasse 1, 85748 Garching, Germany\\
$^{3}$~Center for Space and Habitability, Universit\"{a}t Bern, 3012 Bern, Switzerland\\
$^{4}$~NCCR PlanetS - Universit\"{a}t Bern, Physikalisches Institut, Universit\"{a}t Bern, 3012 Bern, Switzerland\\
$^{5}$~Heidelberg University, Center for Astronomy, Institute for Theoretical Astrophysics, Albert-Ueberle-Strasse 2, 69120 Heidelberg, Germany\\
$^{6}$~European Southern Observatory, Karl-Schwarzschild-Strasse 2, 85748 Garching, Germany}

\begin{document}

\date{Accepted xxx.  Received xxx; in original form xxx}

\pagerange{\pageref{firstpage}--\pageref{lastpage}} \pubyear{2016}

\maketitle

\label{firstpage}

\begin{abstract}
Low-mass protostars are the extrasolar analogues of the natal Solar System. Sophisticated physicochemical models are used to simulate the formation of two protoplanetary discs from the initial prestellar phase, one dominated by viscous spreading and the other by pure infall. The results show that the volatile prestellar fingerprint is modified by the chemistry en route into the disc. This holds relatively independent of initial abundances and chemical parameters: physical conditions are more important. The amount of CO$_{2}$ increases via the grain-surface reaction of OH with CO, which is enhanced by photodissociation of H$_{2}$O ice. Complex organic molecules are produced during transport through the envelope at the expense of CH$_{3}$OH ice. Their abundances can be comparable to that of methanol ice (few \% of water ice) at large disc radii ($R>30$~AU). Current Class II disc models may be underestimating the complex organic content. Planet population synthesis models may underestimate the amount of CO$_{2}$ and overestimate CH$_{3}$OH ices in planetesimals by disregarding chemical processing between the cloud and disc phases. The overall C/O and C/N ratios differ between the gas and solid phases. The two ice ratios show little variation beyond the inner $10$~AU and both are nearly solar in the case of pure infall, but both are sub-solar when viscous spreading dominates. Chemistry in the protostellar envelope en route to the protoplanetary disc sets the initial volatile and prebiotically-significant content of icy planetesimals and cometary bodies. Comets are thus potentially reflecting the provenances of the midplane ices in the Solar Nebula.
\end{abstract}


\section{Introduction}
\label{intro}

Protoplanetary discs encircling young protostars are the birth locations of future mature planetary systems (see \citealt{Johansen2014} for a review). Past observations have suggested that dust growth from micron sizes may begin as early as the prestellar core stage and dust grains may reach millimetre dimensions during infall in the envelopes of protostars (\citealt{Pagani2010, Miotello2014, Jones2016, Ysard2016}, see \citealt{Testi2014} for a review). Recent ALMA data have shown strongly defined ring structures in the disc around the Class I--II protostar HL Tau, which is thought to be younger than $\leq 1 - 2$~Myr and still embedded in a large envelope \citep{ALMA2015}. Models have suggested that these rings may be caused by several planets of at least $0.2$~M$_{\text{J}}$ in mass clearing gaps in the dust distribution \citep{Dipierro2015, Pinte2016, Dong2015}. An alternative hypothesis is that the contrast reflects enhanced dust growth to centimeter sizes behind various snowlines, leading to a change in opacity in the emitting dust \citep{KeZhang2015}. This is potentially aided by sintering close to the snowlines of the volatiles \citep{Okuzumi2016}. These findings are pushing the onset of planetesimal formation much earlier along the star-disc evolutionary sequence than previously thought, perhaps even as early as the embedded phase.

At the same time, the exoplanet community has demonstrated the diverse outcomes of planet formation (see, e.g., www.exoplanet.eu, \citealt{Schneider2011}). Planet population synthesis modellers have carried out pioneering work in linking protoplanetary disc theory with the final architecture of planetary systems. Such models take the key physical processes across all these evolutionary stages into account, and with large sets of initial conditions, to make statistical predictions on the exoplanet population (see \citealt{Benz2014} for a review). It has been postulated that planetary atmospheres form initially via pebble accretion and heating from this assembly prior to runaway gas accretion \citep{InabaIkoma2003, OrmelKlahr2010, Bitsch2015}. Based on the newest results from the star formation community, it may be necessary to use the initial conditions from the earlier embedded phase rather than the classical T Tauri (Class II) stage.

Typically, the constituents of planetesimals that feed protoplanets are grouped into volatiles (H, O, C, N and S containing molecules) and refractories/rocks (minerals/inorganic refractories and complex organic refractory components such as PAHs and macromolecular complex organic matter). For regions of a protoplanetary disc where most volatiles are frozen out as solids, an ice/rock mass ratio of $\sim 2 - 4$ is suggested \citep{Pontoppidan2014}. Various instruments aboard the \textit{Rosetta} mission are attempting to constrain the the icy and dusty contents of comet 67P/Churyumov-–Gerasimenko to an unprecedented precision. The RSI experiment indicates a ice/dust mass of ratio of $\sim 4$ based on the gravity field of the comet \citep{Patzold2016}. CONSERT measurements of an ice/dust volumetric ratio of $\sim 0.4-2.5$ \citep{Kofman2015} imply a ice/dust mass ratio of $\sim 0.1-0.9$, assuming an average dust density of $2~600$~kg~m$^{-3}$ and an ice density of $940$~kg~m$^{-3}$ as in \citet{Patzold2016}. These ratios mean that the solid-phase chemical composition is a pivotal parameter for the subsequent protoplanet makeup. The density is highest in the midplane of a protoplanetary disc, thus harbouring the bulk of the mass and implying that predominantly the ices in that region will likely shape the chemical composition of the atmospheres of the exoplanets that are observed today.

The chemical composition of discs in the embedded phase has been probed by models of \citet{Visser2009}, \citet{Visser2011}, \citet{Drozdovskaya2014}, \citet{Harsono2015}. Observational evidence for them is scarce, because high spatial resolution ($\lesssim 30$~AU) is necessary to get spatial information for a small ($\sim 100$~AU) target,whose emission is easily overwhelmed by that of the more massive envelope, making it hard to constrain their physical parameters. Only recently, observations have shown embedded Keplerian discs of $\sim 50 - 300$~AU in radius in embedded protostars \citep{Tobin2012, Tobin2013, Tobin2015, Brinch2013, Murillo2013, Sakai2014, Harsono2014, Chou2014}. Instead, significant modelling effort has focused on discs in the Class II/III stages of star formation (see, e.g., \citet{HenningSemenov2013} and table~$1$ in \citealt{Woitke2016} and the references therein). The chemistry in such Class II discs has been modelled by means of gas-grain models with grain-surface reactions, e.g., \citet{Willacy2006, Semenov2006, Hersant2009, Semenov2010, Henning2010, Walsh2010, Walsh2012, Vasyunin2011, Akimkin2013}. Only recently complex organics (loosely defined in both chemistry and astronomy as large, $\geq 6$ atoms, carbon-containing species, \citealt{HerbstvanDishoeck2009}) have also been considered by \citet{SemenovWiebe2011}, \citet{Furuya2014}, \citet{Walsh2014a}, \citet{Walsh2014b}, see also \citet{HenningSemenov2013} for a review. Meanwhile, such molecules have been frequently observed towards protostars and in Solar System bodies \citep{HerbstvanDishoeck2009}.

The aim of this work is to compute the chemical composition of planetesimals and cometary materials in the midplanes of protoplanetary discs in the embedded phase of star formation. This is done using state-of-the-art physicochemical models that are dynamic in nature and include chemical kinetics, building upon the work in \citet{Drozdovskaya2014}, which studied the discs as a whole. Main volatiles that are expected to dominate the icy component of plantesimals in the midplanes of protoplanetary discs are considered first. Trace complex organic species are also analysed in this work, since they are potentially the earliest precursors to prebiotic molecules. The dynamic nature of the model allows the study of the chemical effects stemming from the transport from the prestellar cloud through the envelope and into the disc. The key model features are highlighted in Section~\ref{model}. The results are shown in Section~\ref{results} and are discussed in Section~\ref{discussion}, including all the derived implications for the population synthesis models and the Solar System community. The conclusions are presented in Section~\ref{conclusions}.

\section{Models}
\label{model}

\ctable[
 star = 1,
 caption = Molecular abundances at the end of the prestellar phase\tmark and their binding energies$^{b}$.,
 label = tbl:precollapse
 ]{@{\extracolsep{\fill}}llllll}{
 \tnote{The constant physical conditions of the prestellar phase are: $n_{\text{H}}=4 \times 10^{4}$~cm$^{-3}$, $T_{\text{dust}}=10$~K and only cosmic ray-induced FUV photons, $\zeta = 5 \times 10^{-17}$~s$^{-1}$. The duration is taken to be $3 \times 10^{5}$~yr.}
 \tnote[b]{The differences between this table and table~2 of \citealt{Drozdovskaya2015} are due to this work including the glycolaldehyde formation via glyoxal, reaction-diffusion competition, a coverage factor for dust grains (limiting the number of reactive layers) and a higher barrier thickness between grain surface sites.}
 \tnote[c]{\citet{Fraser2001} - measurement for pure water ice in the thin films ($\ll 50$ monolayers) regime (within errors of measurements of \citet{Collings2015} for pure water ice in the multilayer regime)}
 \tnote[d]{\citet{Collings2004, GarrodHerbst2006} - measurement for CO on H$_{2}$O ice (to account for the deep trapping in the inner layers seen with multilayer models of \citep{Taquet2012} and for the trapping in pores of amorphous water ice seen in the experiments of \citet{Collings2003a}); and measurements for NH$_{3}$ and H$_{2}$S on H$_{2}$O ice}
 \tnote[e]{\citet{Edridge2010} - measured average for pure carbon dioxide ice in the multilayer regime}
 \tnote[f]{\citet{Herrero2010} - measurement for CH$_{4}$ on H$_{2}$O ice (to account for the diffusion of methane molecules into the micropores of amorphous solid water (ASW) seen in these experiments)}
 \tnote[g]{\citet{BrownBolina2007} - measurement for pure methanol ice in the multilayer regime}
 \tnote[h]{\citet{Oberg2005} - measurement for pure N$_{2}$ ice in the multilayer regime (within errors of measurements of \citet{Bisschop2006}, \citet{Collings2015}, and \citet{Fayolle2016} for $^{15}$N$_{2}$, for pure ices in the multilayer regime)}
 \tnote[i]{\citet{GarrodHerbst2006}}
 \tnote[j]{\citet{Oberg2009} - measurements for pure ices in the multilayer regime}
 \tnote[k]{\citet{Garrod2013a}}
 }{
 \hline
 Species & Name & $n \left( \text{X}_{\text{gas}} \right) / n_{\text{H}}$ & $n \left( \text{X}_{\text{ice}} \right) / n_{\text{H}}$ & $n \left( \text{X}_{\text{ice}} \right) / n \left( \text{H}_{2}\text{O}_{\text{ice}} \right)$ & $E_{\rm des}({\rm X})$ (K) \T\B\\
 \hline
 H$_{2}$O          & water            & $7.7\times10^{-8}$  & $1.9\times10^{-4}$  & $1.0\times10^{0}$ & $5773^{c}$ \T\\
 CO                & carbon monoxide  & $1.4\times10^{-5}$  & $5.9\times10^{-5}$  & $3.1\times10^{-1}$ & $1150^{d}$ \\
 CO$_{2}$          & carbon dioxide   & $7.9\times10^{-8}$  & $1.9\times10^{-5}$  & $1.0\times10^{-1}$ & $2990^{e}$ \\
 NH$_{3}$          & ammonia          & $2.1\times10^{-7}$  & $4.8\times10^{-6}$  & $2.5\times10^{-2}$ & $5534^{d}$ \\
 CH$_{4}$          & methane          & $1.2\times10^{-7}$  & $1.5\times10^{-5}$  & $7.9\times10^{-2}$ & $1090^{f}$ \\
 CH$_{3}$OH        & methanol         & $2.0\times10^{-10}$ & $3.7\times10^{-6}$  & $1.9\times10^{-2}$ & $4930^{g}$ \\
 H$_{2}$S          & hydrogen sulfide & $5.3\times10^{-10}$ & $8.1\times10^{-9}$  & $4.3\times10^{-5}$ & $2743^{d}$ \\
 N$_{2}$           & nitrogen         & $1.3\times10^{-5}$  & $1.1\times10^{-5}$  & $5.8\times10^{-2}$ & $~~790^{h}$ \\
 H$_{2}$CO         & formaldehyde     & $9.7\times10^{-9}$  & $2.3\times10^{-6}$  & $1.2\times10^{-2}$ & $2050^{i}$ \\
 CH$_{2}$CO        & ketene           & $4.6\times10^{-10}$ & $8.0\times10^{-8}$  & $4.2\times10^{-4}$ & $2200^{i}$ \\
 HCOOH             & formic acid      & $2.4\times10^{-11}$ & $4.9\times10^{-11}$ & $2.6\times10^{-7}$ & $5000^{j}$ \\
 HCOOCH$_{3}$      & methyl formate   & $1.1\times10^{-10}$ & $6.4\times10^{-14}$ & $3.4\times10^{-10}$ & $4000^{j}$ \\
 CH$_{3}$CHO       & acetaldehyde     & $1.0\times10^{-9}$  & $1.5\times10^{-6}$  & $7.9\times10^{-3}$ & $3800^{j}$ \\
 CH$_{3}$OCH$_{3}$ & dimethyl ether   & $2.4\times10^{-11}$ & $9.2\times10^{-8}$  & $4.8\times10^{-4}$ & $3300^{j}$ \\
 C$_{2}$H$_{5}$OH  & ethanol          & $2.2\times10^{-11}$ & $4.7\times10^{-8}$  & $2.5\times10^{-4}$ & $5200^{j}$ \\
 CH$_{3}$COOH      & acetic acid      & $2.2\times10^{-17}$ & $9.7\times10^{-14}$ & $5.1\times10^{-10}$ & $6300^{j}$ \\
 HOCH$_{2}$CHO     & glycolaldehyde   & $2.2\times10^{-12}$ & $5.3\times10^{-9}$  & $2.8\times10^{-5}$ & $6680^{k}$ \B\\
 \hline}

Models simulating the formation of protostars typically start from the so-called prestellar core/cloud phase, which undergoes spherical collapse. Cloud rotation leads to the formation of a protoplanetary disc and associated bipolar outflows due to the conservation of angular momentum.  This work and the preceding publication by \citet{Drozdovskaya2014} use the axisymmetric, 2D semi-analytic physical model developed by \citet{Visser2009} (and \citealt{VisserDullemond2010, Visser2011, Harsono2013}), which describes the density and velocity of the collapsing material. Wavelength-dependent radiative transfer calculations are performed with \textsc{RADMC-3D}\footnote[2]{\url{http://www.ita.uni-heidelberg.de/~dullemond/software/radmc-3d/}} to compute the dust temperature and far-ultraviolet radiation field (FUV; $912 - 2066$~$\text{\AA}$; $6.0 - 13.6$~eV). Thereafter, the physical parameters are coupled with a large chemical network in the framework of a two-phase (gas and solid) model solved using the rate-equation method, including grain-surface chemistry and the chemistry of several families of complex organic molecules \citep{McElroy2013, GarrodHerbst2006, Walsh2014b, Walsh2014a, Walsh2015}. Initial atomic abundances are evolved under constant prestellar cloud conditions ($n_{\text{H}} = 4 \times 10^{4}$~cm$^{-3}$, $T_{\text{dust}} = 10$~K and cosmic ray-interaction with H$_{2}$ as the only source of FUV photons; assuming a cosmic ray ionisation rate $\zeta_{0} = 5 \times 10^{-17}$~s$^{-1}$) for $3 \times 10^{5}$~yr (table~$2$ of \citealt{Drozdovskaya2014}) to obtain initial molecular abundances at the onset of collapse (Table~\ref{tbl:precollapse}). Chemical abundances are computed along evolutionary infall trajectories that are obtained from the physical model with path-dependent physical conditions. The collapse of the system takes an additional $2.46 \times 10^{5}$~yr, which corresponds to the time it takes for the envelope to accrete on to the star-disc system for the initial physical conditions chosen for this work.

\begin{figure}
 \centering
 \includegraphics[width=0.45\textwidth,keepaspectratio]{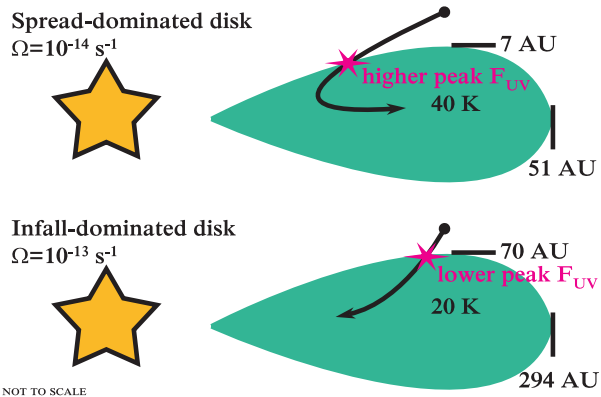}
 \caption{An illustration of the key parameters (such as the midplane dust temperature, outer radius and predominant parcel motion) of the two discs that are central to this work and \citet{Drozdovskaya2014}.}
 \label{fgr:physmodel}
\end{figure}

In this paper, as in \citet{Drozdovskaya2014}, two discs are studied. One is the so-called spread-dominated case, with a lower initial cloud rotation rate, in which the disc primarily grows via viscous spreading and has a final outer radius of $\sim 50$~AU. The second is the infall-dominated case, in which the disc is assembled predominantly via pure infall of material and has a final outer radius of $\sim 300$~AU (table~$1$ of \citealt{Drozdovskaya2014}). The reason for studying both discs is that their unique large-scale dynamics result in different dust temperature and FUV flux profiles along the trajectories, i.e., they have different physical histories for the protoplanetary disc material (this is pictorially illustrated in Fig.~\ref{fgr:physmodel}). At $\sim 40$~K, the smaller spread-dominated disc is warmer than the more massive infall-dominated disc, which is $\sim 20$~K close to the midplane (middle panels of figs.~4 and~5 of \citealt{Drozdovskaya2014}). A larger disc mass hinders passive heating by reprocessed stellar radiation. These models exclude viscous heating and its effects, such as those discussed in \citep{Harsono2015}; however, this is most important only very close to the star. These models do not consider magnetic field effects and dust growth, as discussed in, e.g., \citep{Zhao2016} and \citep{PaulyGarrod2016}.

Once collapse is initiated, the central protostar is the dominant source of FUV photons. Its luminosity, which varies with time and peaks at $\sim 11$~L$_{\sun}$ for the spread-dominated case (fig.~$2$ of \citealt{Drozdovskaya2014}), is given by the sum of the accretion luminosity \citep{AdamsShu1986} and the photospheric luminosity (due to gravitational contraction and deuterium burning; \citealt{DAntonaMazzitelli1994, Visser2009}). Excess UV observed for classical T Tauri stars (effective temperatures of $\sim 10~000$~K) as a result of boundary layer accretion from the disk onto the star \citep{Bertout1988} is not accounted for in this work. External sources of FUV are excluded, since such young objects are expected to be deeply embedded in molecular clouds and hence highly shielded. The maximal FUV flux experienced by a parcel is set by the $1/r^{2}$ dependence of the flux and the corresponding decrease in the attenuating column as the parcel moves closer to the star. Once the parcel enters the disk, the attenuating column increases substantially, leading to an almost instantaneous reduction in FUV flux. The material entering the spread-dominated disc experiences higher peak FUV irradiation, because many of the parcels pass closer to the protostar prior to disc entry than in the infall-dominated case. The smaller disc also exerts less attenuation of cosmic rays, thus there are less internally produced cosmic ray-induced FUV photons in the infall-dominated disc. Both, the dust temperature and FUV flux, are crucial parameters for chemical grain-surface reactions controlling the mobility and availability of radicals.

All further details on the physical and chemical model and the radiative transfer setup can be found in \citet{Drozdovskaya2014, Drozdovskaya2015}. In this work, a different set of parcels is considered - more than $100$ for each disc, all with final positions at $z = 0.01$~AU and $R$ in the $\sim 1-50$~AU range for the spread-dominated disc ($\sim 1-300$~AU range for the infall-dominated disc) to better sample the two midplanes. For the spread-dominated case at $R=5.2$~AU, $\sim 90$ per cent of the mass lies below $1$~AU in height, while the disc surface extends up to $\sim 2$~AU. This work includes the experimentally suggested glycolaldehyde formation via glyoxal (\citealt{Fedoseev2015}, as described in section 3.4.3 of \citealt{Drozdovskaya2015}). Furthermore, diffusion-reaction competition is now accounted for via the formulation of \citet{GarrodPauly2011} (equation~6). This enhances the probability of reactions with activation barriers ($E_{\text{A}}$) when thermal diffusion rates are low. A larger $1.5~\text{\AA}$ barrier thickness ($a$) between grain surface sites is assumed to decrease the efficiency of quantum tunnelling reactions involving atomic and molecular hydrogen. The effects of varying the barrier to quantum tunneling are investigated in Section~\ref{barriers}. Furthermore, a `coverage factor' of $2$ ice monolayers has been imposed. This parameter implies that grain-surface associations can occur only in the two upper most monolayers. However, in a two-phase model the surface is not distinguished from the bulk in terms of composition; therefore, the composition of these two chemically active layers is averaged together with that of the non-chemically active bulk. Finally, since gas column densities at positions along the midplane mostly exceed $100$~g~cm$^{-2}$, cosmic ray attenuation is now accounted for according to
\begin{equation}
 \zeta = \zeta_{0}~\text{exp}\left( \frac{-\Sigma_{\left( R,z \right)}}{96~\text{g cm}^{-2}} \right),
\end{equation}
where $\Sigma_{\left( R,z \right)}$ is the column density between the position of interest $\left( R,z \right)$ and the disc surface at that time ($\Sigma_{\left( R,z \right)} = \mu m_{u} \int ^{\text{surface}}_{z} n_{\text{gas}} dz$, where $\mu=2.3$ is the mean molecular mass of the gas, $m_{u}$ is the atomic mass unit, $n_{\text{gas}}$ is the gas particle number density; fully molecular gas is assumed for simplicity, i.e., $n_{\text{gas}}=n_{\text{H}_{2}}$ and $n_{\text{H}}=2n_{\text{H}_{2}}$). A lower limit of $7.3 \times 10^{-19}$~s$^{-1}$ is imposed, which stems from $^{26}$Al radionuclide decay \citep{UmebayashiNakano1981, Cleeves2013}.

\subsection{Initial conditions: `hot' disc start versus `cold' cloud start scenarios}
\label{coldhot}
The typical initial assumption of the solar system community and population synthesis models is that the initial conditions of the Solar Nebula are hot, i.e., all disc material (whether volatile or refractory) is in the gas phase. The disc is thought to be small, dense and hot (heated by viscous dissipation, initially), and subsequently to expand in radius and cool. Typically, to bypass this evolution, the initial temperature is simply assumed high ($\gg 300$~K) out to at least $30$~AU. For planet population synthesis statistics, disc models provide radial pressure, temperature and surface density profiles \citep{Alibert2005, Alibert2013, Marboeuf2014a, Marboeuf2014b, Thiabaud2014, Thiabaud2015}. Individual volatiles are then allowed to condense instantaneously at their respective temperature-pressure of condensation (and also be trapped in clathrates, so-called water cages that form from crystalline water ice, the importance of which is still under debate, e.g., \citealt{LunineStevenson1985, Mousis2012, Thurmer2015}), assuming thermodynamic equilibrium. This forms the ice and rock components of small grains as a function of disc radius ($R$). It is assumed that the composition of $300$~m planetesimals at a certain $R$ in the disc is identical to that of the small grain composition. The population synthesis model then describes how $10$ planetary Lunar-sized embryos (of the same composition as the planetesimals at a certain $R$) grow through planetesimal accretion during migration, while the disc evolves and disperses by photoevaporation. By running a very large number of disc models with varying initial conditions, general trends in planet populations are studied \citep{Marboeuf2014a, Marboeuf2014b, Thiabaud2014}. No chemistry occurs in these models; variations in the ice composition are set by the locations of snowlines (condensation fronts) at the start ($t=0$). In contrast, the chemical models of the star formation community include a kinetic chemical network and many, if not all, of the possible chemical processes at play.

The initially hot Solar Nebula (`hot' disc start scenario) versus the cold prestellar phase and warm protoplanetary disc (`cold' cloud start scenario) is one of the major differences between the approaches of the Solar System and star plus disc formation communities. Under the `cold' start scenario, a large icy reservoir is initially in place, which is not necessarily mimicked by the gas. Meanwhile, for the `hot' start, only gases are available at the start. It remains unclear if potentially both scenarios are valid in different regions of the disc (see the review of \citealt{Davis2014}). Calcium–-aluminium-rich inclusions (CAIs) are typically presented as evidence for the `hot' scenario, since their production occurs at temperatures $>1300$~K. CAIs are found in meteorites, which mostly originate from the asteroids in the inner Solar System, but have also been detected in comet $81$P/Wild~$2$ with the Stardust mission \citep{Brownlee2006}. High temperature-processing is also suggested by the presence of crystalline silicates in protoplanetary discs and comets, which can only form by annealing at temperatures of $800 - 1000$~K or via condensation upon cooling below $\sim 1200-1500$~K. However, all observations of prestellar and protostellar sources point towards the `cold' and `warm' set of conditions. Localized temperature enhancement could occur via accretion shocks at the disc surface or via episodic accretion with subsequent spreading of crystalline silicates by radial or vertical large-range mixing throughout the protoplanetary disc (e.g., \citealt{VisserDullemond2010, Ciesla2011}). It is still debated whether either of these processes can reach high enough temperatures, but if so, strongly heated materials would be available at the largest disc radii for incorporation into distant larger bodies.

\begin{figure*}
 \centering
 \includegraphics[width=0.75\textwidth,keepaspectratio]{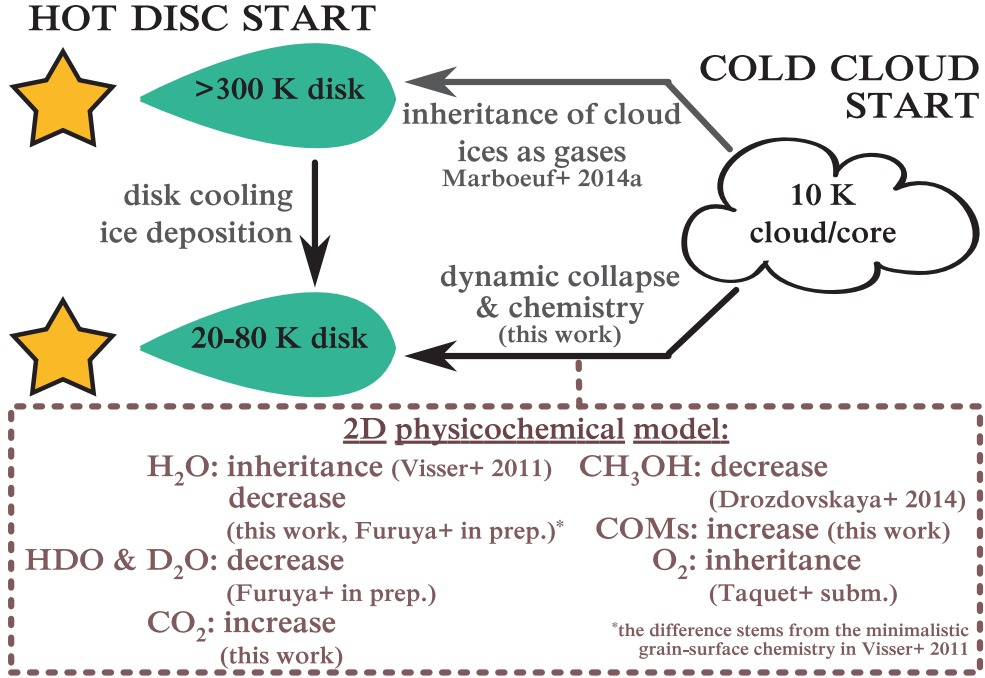}
 \caption{An illustration of the `hot' and `cold' start perspectives. The results obtained with this physical model in various publications, including this work, are also summarized.}
 \label{fgr:scenarios}
\end{figure*}

The models of \citet{Marboeuf2014b, Marboeuf2014a} and \citet{Thiabaud2014, Thiabaud2015} impose initial gas-phase abundances based on observations of prestellar cores and protostellar sources, assuming complete inheritance without modification of the prestellar icy composition by discs upon formation. In other words, they are a hybrid of the two scenarios just discussed. This is visually summarized in Fig.~\ref{fgr:scenarios}. This work tests the simplified approach of phase diagrams under the assumption of complete prestellar inheritance to obtain the volatile composition of discs by comparing with the output obtained from a full kinetic chemical model, which also takes dynamic collapse into account. The focus lies on the disk beyond the inner few AU.

\section{Results}
\label{results}

\begin{figure}
 \centering
 \includegraphics[width=0.45\textwidth,keepaspectratio]{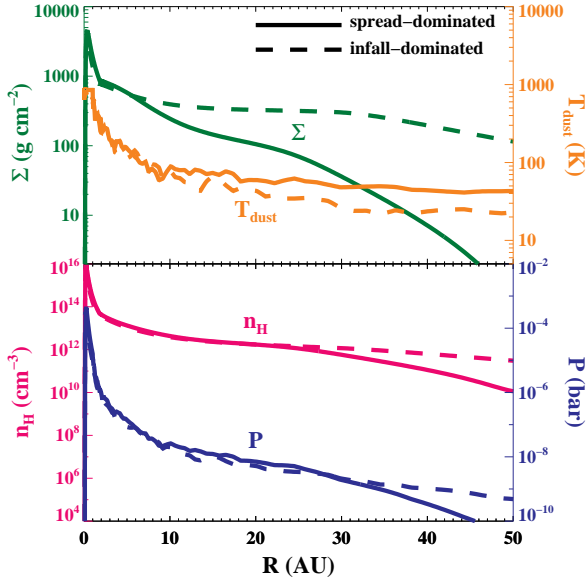}
 \caption{The midplane ($z \approx 0.01$~AU) physical conditions of the spread-(solid lines) and infall-dominated (dashed lines) discs at the end of the simulation ($2.46 \times 10^{5}$~yr after the precollapse phase): dust temperature ($T_{\text{dust}}$), H nuclei number density ($n_{\text{H}}$), surface density ($\Sigma$) and pressure ($P$) as a function of disc radius ($R$). The infall-dominated disc extends out to $300$~AU, only the inner $50$~AU is shown.}
 \label{fgr:midplanes}
\end{figure}

In Fig.~\ref{fgr:midplanes}, the midplane conditions for both the spread- and infall-dominated discs are shown (disc slice at $z \approx 0.01$~AU). These are the physical conditions that the parcels reach at the final timestep of the evolution, depending on their final radial coordinate.

The inner few AU of each midplane, close to the protostar, has the highest final dust temperature ($T_{\text{dust}}>200$~K) and density ($n_{\text{H}}>10^{14}$~cm$^{-3}$). The smaller spread-dominated disc tends to be warmer ($\sim40$~K for $R>25$~AU) than the more massive infall-dominated disc ($\sim20$~K for $R>30$~AU) due to more efficient passive heating. For $R \gtrsim 25$~AU, the density, surface density ($\Sigma=2 \mu m_{u} \int ^{\text{surface}}_{0.01~\text{AU}} n_{\text{gas}} dz$, where the initial factor of $2$ accounts for the full vertical extent of the disc) and pressure ($P=k_{\text{B}}n_{\text{gas}}T_{\text{gas}}$, where $k_{\text{B}}$ is the Boltzmann constant and assuming $T_{\text{gas}} = T_{\text{dust}}$) of the midplane of the infall-dominated disc are higher than that of the spread-dominated disc. The surface density typically exceeds $100$~g~cm$^{-2}$ (except for $R>40$~AU for the spread-dominated case) justifying the need for cosmic ray attenuation. The midplane pressure of each disc never exceeds $0.5$~millibar (beyond $5$~AU: $P < 10^{-7}$~bar).

\subsection{Dominant simple ices}
\label{sec:simple}

\begin{figure*}
 \centering
 \begin{subfigure}[b]{0.45\textwidth}
 \includegraphics[width=\textwidth,keepaspectratio]{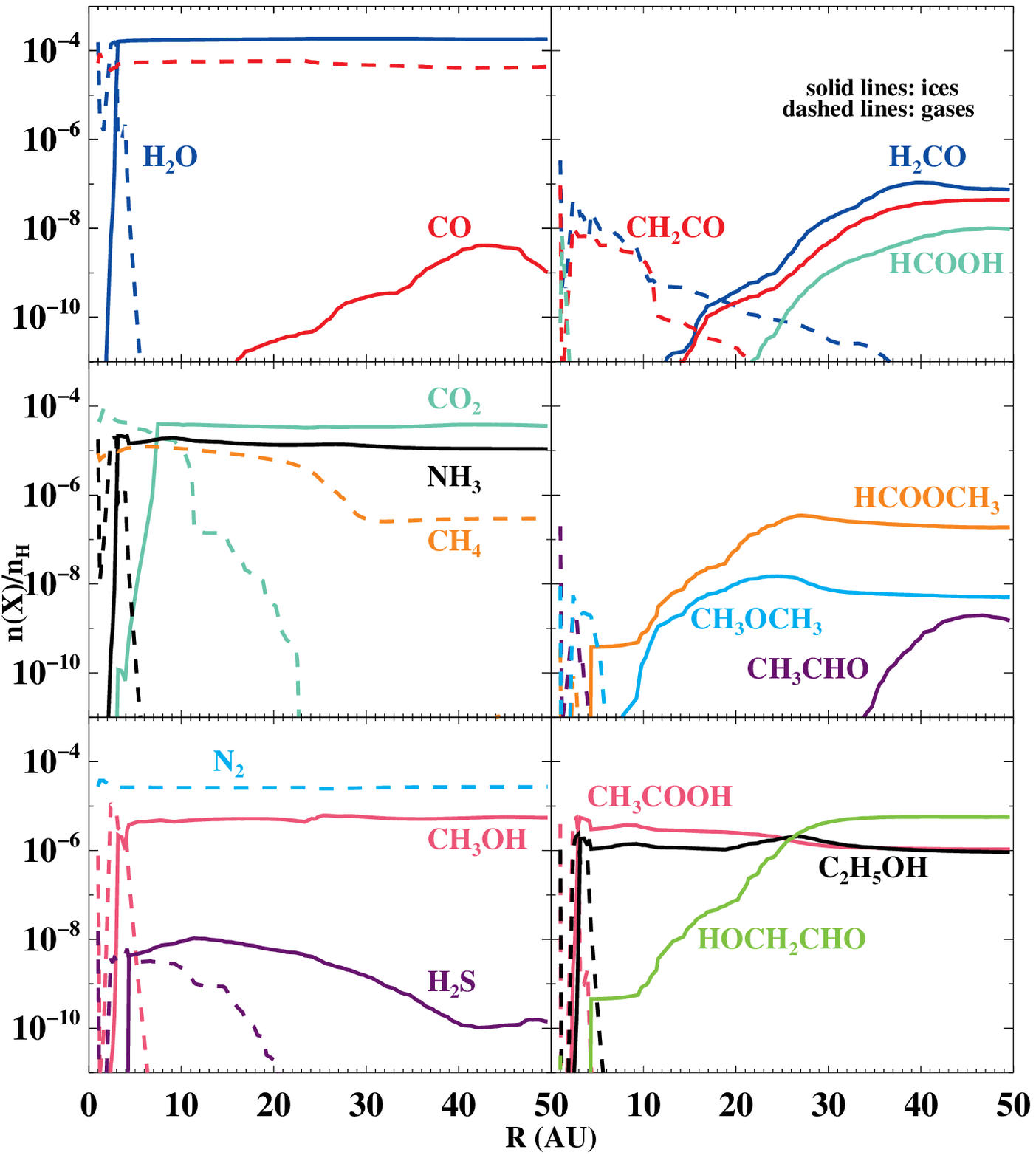}
 \caption{Spread-dominated disc}
 \label{fgr:case3}
 \end{subfigure}
 \begin{subfigure}[b]{0.45\textwidth}
 \includegraphics[width=\textwidth,keepaspectratio]{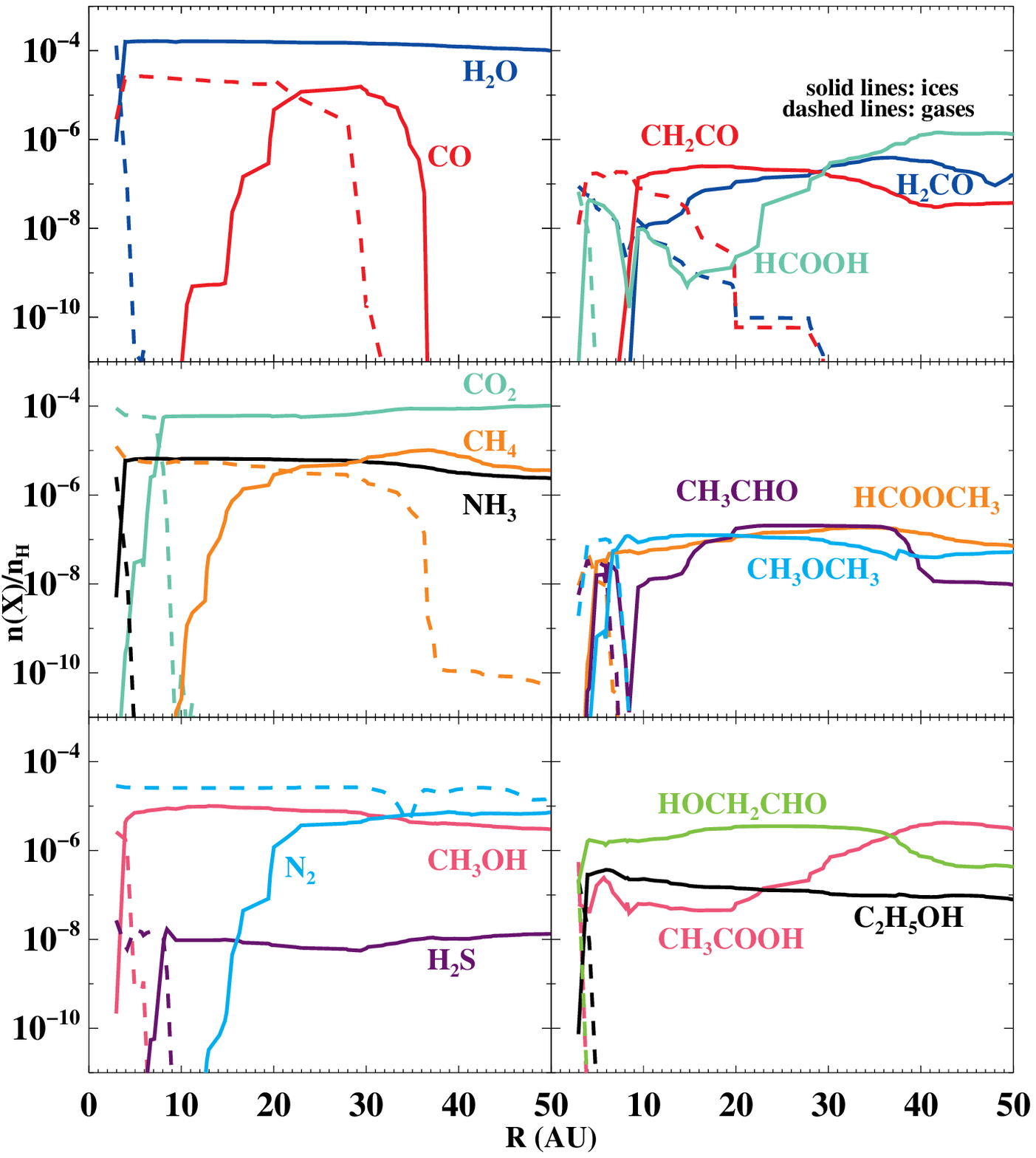}
 \caption{Infall-dominated disc}
 \label{fgr:case7}
 \end{subfigure}
 \caption{The midplane abundances of volatiles (left column of each subfigure) and complex organics (right column of each subfigure) relative to the number of H nuclei ($n_{\text{H}}$) as a function of disc radius ($R$) in the gaseous (dashed lines) and solid (solid lines) phases at the end of the simulation ($2.46 \times 10^{5}$~yr after the onset of collapse). Ice abundances beyond the respective snowlines (Table~\ref{tbl:snowlines}) are tabulated in Appendix~\ref{iceabun}.}
\end{figure*}

\ctable[
 caption = Snowlines$^{l}$ in the midplanes of the spread-dominated and infall-dominated discs at the end of the simulation ($2.46 \times 10^{5}$~yr after the onset of collapse).,
 label = tbl:snowlines
 ]{@{\extracolsep{\fill}}lcc}{
 \tnote[l]{Defined as the point when $n \left( X_{\text{ice}} \right) / n \left( X_{\text{gas}} \right) = 0.5$ and rounded to the nearest AU. Note that these do not always correspond to a snowline in the traditional sense, for example, see the curve of HCOOH in the spread-dominated disc in Fig.~\ref{fgr:case3}, which is set by chemistry.}
 }{
 \hline
 \multirow{2}{*}{Species} & \multicolumn{2}{c}{$R_{\text{snow}}$ (AU)} \T\\
 \cline{2-3}
   & \multicolumn{1}{c}{spread-dom.} & \multicolumn{1}{c}{infall-dom.} \T\B\\
 \hline
 H$_{2}$O          & $~~~~~~3$ & $~~3$\T\\
 CO                & $>50$     & $21$\\
 CO$_{2}$          & $~~~~~~7$ & $~~7$\\
 NH$_{3}$          & $~~~~~~3$ & $~~3$\\
 CH$_{4}$          & $>50$     & $23$\\
 CH$_{3}$OH        & $~~~~~~4$ & $~~3$\\
 H$_{2}$S          & $~~~~~~7$ & $~~8$\\
 N$_{2}$           & $>50$     & $36$\\
 H$_{2}$CO         & $~~~~20$  & $11$\\
 CH$_{2}$CO        & $~~~~~~9$ & $~~8$\\
 HCOOH             & $~~~~~~4$ & $~~3$\\
 HCOOCH$_{3}$      & $~~~~~~5$ & $~~4$\\
 CH$_{3}$CHO       & $~~~~~~7$ & $~~5$\\
 CH$_{3}$OCH$_{3}$ & $~~~~~~7$ & $~~6$\\
 C$_{2}$H$_{5}$OH  & $~~~~~~3$ & $~~3$\\
 CH$_{3}$COOH      & $~~~~~~3$ & $~~3$\\
 HOCH$_{2}$CHO     & $~~~~~~3$ & $~~3$\B\\
 \hline}

The left panels of Figs.~\ref{fgr:case3} and~\ref{fgr:case7} show the radial distribution of simple ices and gases in the inner $50$~AU in an embedded disc for the spread-dominated and infall-dominated cases, respectively, at the final timestep of the evolution ($2.46 \times 10^{5}$~yr since the onset of collapse). A number of snowlines/icelines are seen in both discs, e.g., a steep ice-to-gas transition for water (H$_{2}$O) and a more gradual transition for methane (CH$_{4}$). The sharpness of the transitions is set by the temperature gradient along the midplane of each disc, which is steeper closer to the protostar (Fig.~\ref{fgr:midplanes}). The order of the snowlines is set by the binding energies used in the chemical network (right column of Table~\ref{tbl:precollapse}). For both discs, the water snowline lies around $\sim 3$~AU. The small spread-dominated disc lacks CH$_{4}$ and nitrogen (N$_{2}$) ices, and is low on carbon monoxide (CO) ice; while the infall-dominated disc is rich in all simple ices for $R \gtrsim 25$~AU due to its overall lower temperature. Table~\ref{tbl:snowlines} tabulates all the snowlines in both discs.

\begin{figure*}
 \centering
 \begin{subfigure}[t]{0.3\textwidth}
 \includegraphics[width=\textwidth]{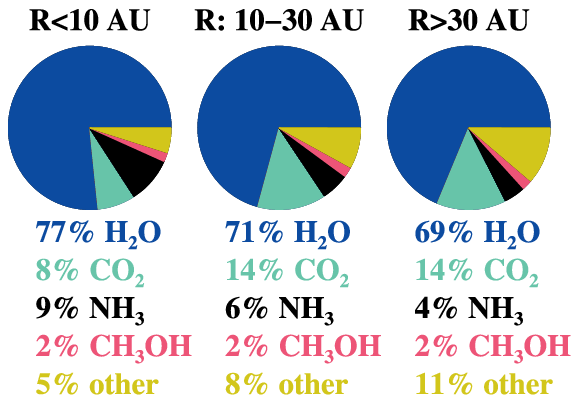}
 \caption{Spread-dominated disc}
 \label{fgr:case3_volpie}
 \end{subfigure}
 \begin{subfigure}[t]{0.3\textwidth}
 \includegraphics[width=\textwidth]{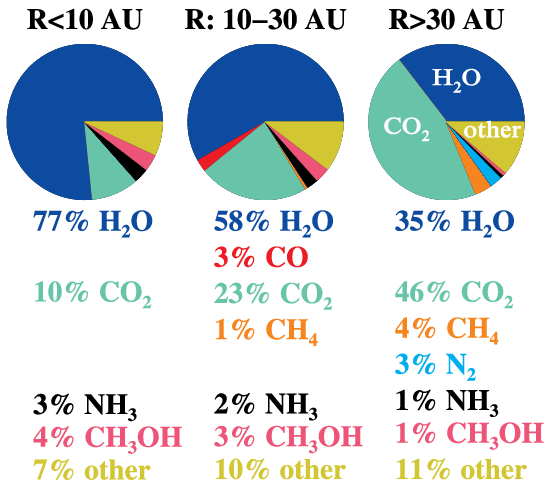}
 \caption{Infall-dominated disc}
 \label{fgr:case7_volpie}
 \end{subfigure}
 \begin{subfigure}[t]{0.3\textwidth}
 \includegraphics[width=\textwidth]{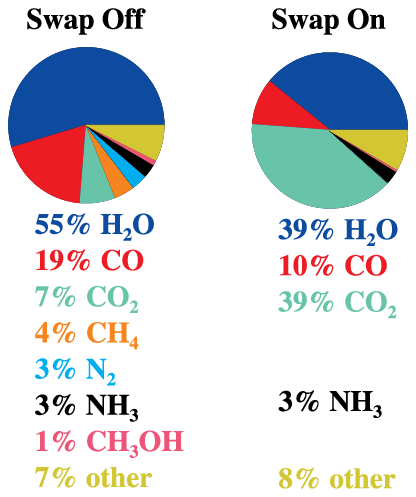}
 \caption{Infall-dominated disc beyond $30$~AU, as obtained with the three-phase chemical model of \citet{Furuya2015}, for bulk and surface combined with and without the swapping mechanism}
 \label{fgr:3phase}
 \end{subfigure}
 \caption{The percentage contribution of volatiles to the total ice content averaged over various radial ranges of the disc midplanes at the end of the simulation ($2.46 \times 10^{5}$~yr after the precollapse phase). Only contributions of at least $\sim 1$\% are shown. The `other' category includes all ices that are not the following eight volatiles: H$_{2}$O, CO, CO$_{2}$, CH$_{3}$OH, CH$_{4}$, N$_{2}$, NH$_{3}$ and H$_{2}$S.}
\end{figure*}

Figs.~\ref{fgr:case3_volpie} and~\ref{fgr:case7_volpie} show pie charts for three radial ranges, illustrating the percentage contribution of each species to the ice mantle. The values are number averages and are calculated according to
\begin{equation}
\frac{n \left( \text{X}_{\text{ice}} \right)}{\sum_{\text{X}} n \left( \text{X}_{\text{ice}} \right)} (\%) = 100\% \times \frac{\sum_{R_{\text{min}}}^{R_{\text{max}}} n \left( \text{X}_{\text{ice}} \right)_{\left( R,z \right)} dR}{\sum_{R_{\text{min}}}^{R_{\text{max}}} \sum_{\text{X}} n \left( \text{X}_{\text{ice}} \right)_{\left( R,z \right)} dR},
\end{equation}
which accounts for radial ranges $dR$ between individual parcels, and where $R_{\text{max}}$ and $R_{\text{min}}$ are the maximum and minimum positions, respectively, of the radial range considered. The pie charts show that typically midplane ice is $\sim 70-80$\% H$_{2}$O, $\sim 10-20$\% carbon dioxide (CO$_{2}$), $1-10$\% ammonia (NH$_{3}$) and $1-4$\% methanol (CH$_{3}$OH). As mentioned above, CO, CH$_{4}$ and N$_{2}$ ices are only available in the cooler infall-dominated disc and are no more than a few per cent of the total ice content. However, this only holds as long as these species are not mixed with H$_{2}$O ice (see the discussion on trapping in Section~\ref{binding}), which would allow these molecules to be retained in the ice at a level of a few per cent. The contributions of various molecules show some radial dependencies, which are a reflection of certain species thermally desorbing at the respective snowlines along the midplane. Other ices, like complex organics and radicals make up no more than $\sim 10$\% of the total.

\begin{figure}
 \centering
 \includegraphics[width=0.45\textwidth,keepaspectratio]{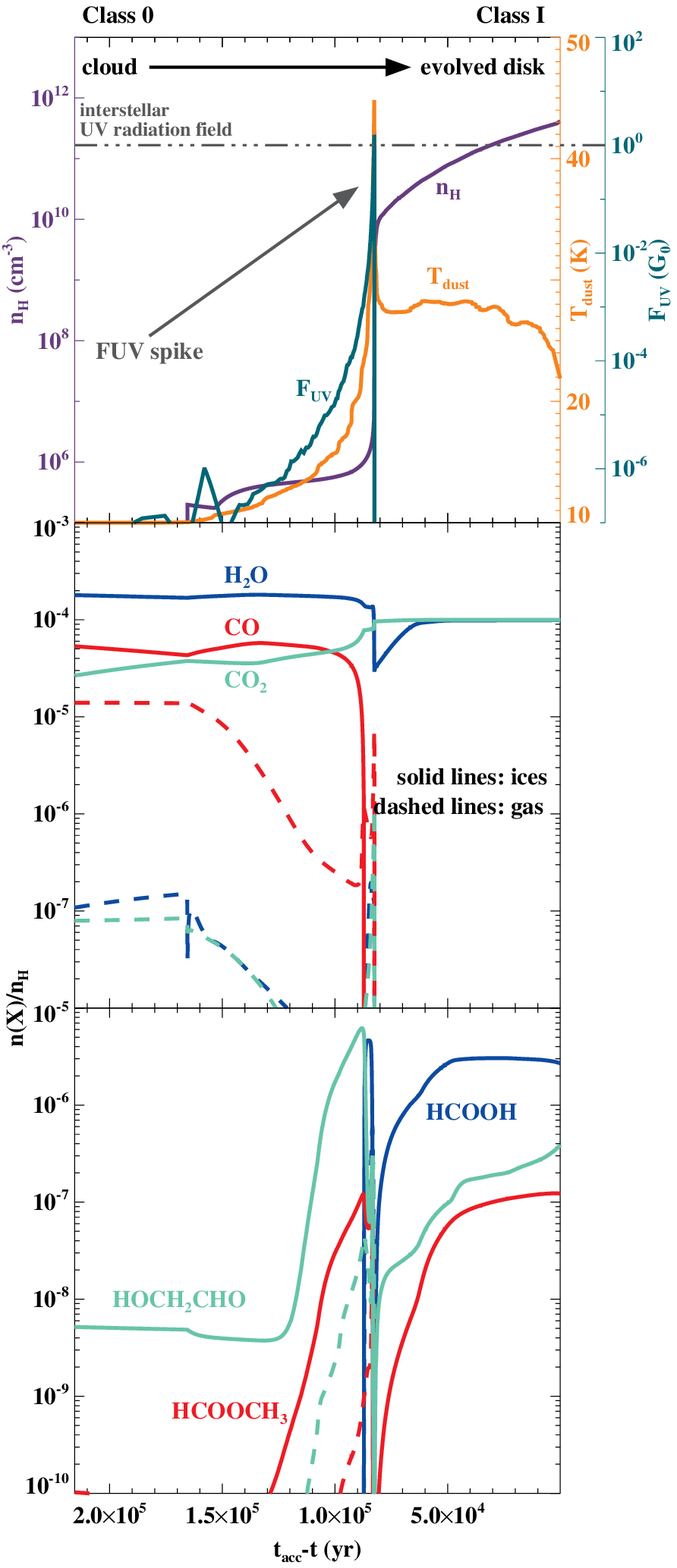}
 \caption{The physical parameters and chemical abundances as a function of time ($t$) for one trajectory (with a final radial position of $\sim 47$~AU) in the case of the infall-dominated disc. The upper panel shows the H nuclei number density ($n_{\text{H}}$), the dust temperature ($T_{\text{dust}}$) and the FUV flux ($F_{\text{UV}}$, in units of $G_{0}$, which is the interstellar FUV radiation field of $1.6 \times 10^{-3}$~erg cm$^{-2}$ s$^{-1}$, \citealt{Habing1968}). The middle panel shows the abundances relative to $n_{\text{H}}$ of water, carbon monoxode and carbon dioxide in the solid and gasesous phases (solid and dashed lines, respectively). The lower panel displays the abundances of formic acid, methyl formate and glycolaldehyde. $t_{\text{acc}}$ is the accretion time of $2.46 \times 10^{5}$~yr in this model run (see \citealt{Drozdovskaya2014} for details). Evolution is from left to right. Note the transformation of CO to CO$_{2}$ ice and the build up of complex organics in the envelope upon the increase of FUV flux.}
 \label{fgr:phys_paper}
\end{figure}

For the infall-dominated disc, in Figs.~\ref{fgr:case7} and~\ref{fgr:case7_volpie} it can be seen that for $R>30$~AU, CO$_{2}$ becomes more abundant than H$_{2}$O ice. This comes from a delicate balance of rehydrogenation of the OH radical (leading to H$_{2}$O) and its reaction with CO on the grain surface (producing CO$_{2}$ with an activation barrier of $400$~K, \citealt{Noble2011}) according to the following scheme.

\schemestart
 H$_{2}$O$_{\text{ice}}$
 \arrow{<=>[\+ h$\nu$][$E_{\text{A}}=0$]}[0,1.5]
 \subscheme{OH} \+ H
 \arrow(@c2--){0}[-90,1.3]
 \subscheme{CO$_{2}$} \+ H
 \arrow(@c3--@c5){->[*{0}\+ CO][$E_{\text{A}}=400$~K]}[-90,1.3]
\schemestop

\noindent This effect comes into play only in the outer regions of the infall-dominated disc, where CO ice is available. The initial photodissociation of H$_{2}$O ice, producing the OH radical, stems from the exposure to stellar FUV photons in the envelope en route to the disc and partially from cosmic ray-induced FUV photons. It appears that the route towards CO$_{2}$ is favoured; and it is efficiently produced at the expense of H$_{2}$O and CO. This is clearly demonstrated when looking at the individual trajectories in detail, as in Fig.~\ref{fgr:phys_paper}, which shows the physical parameters and chemical abundances as a function of time for a parcel that ends up at $R \approx 47$~AU along the midplane. The increase of CO$_{2}$ ice follows the profile of the FUV flux and is maximized once the FUV flux spikes at $\sim G_{0}$ (where $G_{0}$ is the interstellar FUV radiation field of $1.6 \times 10^{-3}$~erg cm$^{-2}$ s$^{-1}$, \citealt{Habing1968}) immediately prior to disc entry. This effect persists under a static assumption (Section~\ref{static}), but to a lesser degree, thus displaying the subdominant importance of cosmic ray-induced FUV photons towards the H$_{2}$O ice photodissociation process. Such efficient CO$_{2}$ production was already noticed in the molecular layer at intermediate heights of Class II protoplanetary disc models of \citet{Walsh2014a} (which also includes a discussion on the uncertainties in this reaction) and in the relatively cool disc around an M dwarf star in \citet{Walsh2015}.

\subsection{Sensitivity of results to chemical parameters}

\subsubsection{Diffusion and quantum tunneling barriers}
\label{barriers}
To verify the robustness of the predicted CO$_{2}$ ice-rich zone in the infall-dominated disc, additional parameter tests were run to test the sensitivity of our results to the assumed grain-surface parameters. In the fiducial setup of our models, the diffusion barrier ($E_{\text{diff}}$) is related to the binding energy ($E_{\text{des}}$, also called the desorption energy) via $E_{\text{diff}} = 0.3 E_{\text{des}}$ \citep{HasegawaHerbstLeung1992}. The ratio between the two energies is uncertain. Off-lattice Monte Carlo simulations predict a value of $\sim 0.3-0.4$ for CO and CO$_{2}$ on amorphous water ice \citep{KarssemeijerCuppen2014, Karssemeijer2014}, while experiments suggest a range of $\sim 0.5-0.9$ for O and N on amorphous water ice \citep{Minissale2016a}; however, it is not excluded that this ratio may differ between molecules and atoms. Astrochemical models typically employ values in the $\sim 0.3-0.8$ range (e.g., \citet{VasyuninHerbst2013a}. For the first test, the factor between the two energies is set to $0.5$ to increase the diffusion barrier, thereby impeding radical-radical associations on grain surfaces and limiting the efficiency of the CO+OH surface reaction. For the second test, the fiducial barrier thickness of $1.5$~$\text{\AA}$ is decreased, leading to a thinner barrier for quantum tunneling of H and H$_{2}$ between grain surface sites and, thus, the promotion of the OH+H surface reaction.


\ctable[
 caption = Abundances ratios of selected solid species at the end of the simulation ($2.46 \times 10^{5}$~yr after the precollapse phase) for a parcel in the infall-dominated disc with final $R \approx 50$~AU with three different sets of parameters in the chemical model.,
 label = tbl:param
 ]{@{\extracolsep{\fill}}llll}{
 \tnote[g]{In the fiducial setup, $E_{\text{diff}} = 0.3 E_{\text{des}}$ and $a = 1.5~\text{\AA}$.}
 }{
 \hline
 Species           & Fiducial$^{g}$     & $E_{\text{diff}} = 0.5 E_{\text{des}}$ & $a = 1~\text{\AA}$ \T\B\\
 \hline
 CO$_{2}$/H$_{2}$O & $1.04$             & $1.28$                                 & $2.76$             \T\\
 CO/H$_{2}$O       & $8.3\times10^{-8}$ & $1.3\times10^{-3}$                     & $4.5\times10^{-3}$ \B\\
 \hline}

Both tests were computed for the same parcel with a final radial coordinate $R \approx 50$~AU at the end of the simulation (in the infall-dominated disc; this is not the same parcel as that shown in Fig.~\ref{fgr:phys_paper}). The same initial molecular abundances as computed with fiducial parameters were used for all tests. Table~\ref{tbl:param} displays the final abundance ratios of the key species in relation to grain-surface chemistry of CO$_{2}$. For a higher diffusion barrier, the abundances of CO and radicals such as OH and HCO are higher in the solid phase, since radical mobility is reduced. However, an increase by $\sim 2$ orders of magnitude in HCO abundance leads to more CO$_{2}$ ice formation via the atom-radical O+HCO reaction. For a thinner barrier for quantum tunneling, less atomic H is available on the grain, since it is easier for it to react with other species. This in turn reduces the availability of OH, reducing the amount of CO$_{2}$ ice produced via OH+CO. However, atomic H is not only used to reform H$_{2}$O, but to also hydrogenate other species, thereby reducing the amount of H$_{2}$O ice as well. The net result is that the ice remains CO$_{2}$-rich. Thus such a CO$_{2}$ ice-rich zone appears characteristic of the outer parts of the infall-dominated disc independent of the exact parameters used in the chemical model within the framework of the two-phase model considered here (although they do change the abundances quantatively). In fact, two additional pathways to CO$_{2}$ have recently been experimentally verified, namely CO+O with $E_{\text{A}} = 1000$~K and H$_{2}$CO+O with $E_{\text{A}} = 335$~K \citep{Minissale2013, Minissale2015}. Currently, they are not included in the chemical network, but are expected to positively contribute to the CO$_{2}$ production.

\subsubsection{Binding energies}
\label{binding}
The availability of specific ices in midplanes is set by their initial abundances, the temperature structure of the disc, and also by the binding energies of these species (Table~\ref{tbl:precollapse}). This is why in the warmer spread-dominated disc, CH$_{4}$ and N$_{2}$ ices are absent, and CO ice is low (Section~\ref{sec:simple}, Fig.~\ref{fgr:case3}). The adopted set of binding energies also sets the exact locations of the snowlines. In this work, the binding energies for molecules on amorphous water ice have been adopted for CO, NH$_{3}$, H$_{2}$S and CH$_{4}$. For the larger complex organic species and methanol, the binding energies for pure ices have been used. The binding energies of pure ices tend to be lower than those on H$_{2}$O ice, which is the dominant ice constituent (Fig.~\ref{fgr:case3_volpie}). For example, for pure CO ice in the multilayer regime $E_{\text{des}}=878$~K (\citealt{Collings2015}, which is within errors of measurements of \citet{Oberg2005}, \citet{Bisschop2006}, \citet{Cleeves2014} and \citet{Fayolle2016} for pure ices in the multilayer regime), while for CO on H$_{2}$O ice $E_{\text{des}}=1150$~K (\citealt{Collings2004, GarrodHerbst2006} or $1320$~K, \citealt{Cleeves2014}).

In the laboratory, thermal desorption of CO ice deposited on top of H$_{2}$O ice is a multi-step process \citep{Sandford1988, Ayotte2001, Collings2003a, Collings2003b, Bisschop2006}. At the cold ($10 - 15$~K) temperatures of prestellar cores, CO ice forms on top of the H$_{2}$O ice mantle \citep{Pontoppidan2006, Oberg2011c2d}; and the layers are frozen in place, since the mobility of such large species is minimal at these temperatures. Upon heating, as much as $20\%$ of CO ice can get trapped in amorphous H$_{2}$O ice as it undergoes phase transitions \citep{Collings2003b}. This implies that a single desorption energy is an oversimplification of the process of thermal desorption of CO. \citet{Visser2009} (based on the approach of \citealt{Viti2004}) attempted to account for this effect by considering four flavours of CO with four different desorption energies, which allowed the authors to preserve $\sim 15\%$ of CO ice in a disc at the end of the simulation. Unfortunately, such an approach is not feasible in this work due to a much larger size of the chemical network and the inclusion of grain-surface reactions (thus making reaction-dependent CO-`flavouring' non-trivial). Deep trapping of CO in the inner layers is seen as a natural consequence in the multilayer models \citep{Taquet2012}.


Similar trapping mechanisms have also been seen experimentally for other volatiles, such as N$_{2}$, CH$_{4}$, CO$_{2}$, O$_{2}$, NH$_{3}$; and noble gases, such as Ar, Kr, Xe \citep{Ayotte2001, Notesco2003, Bar-Nun2007, Galvez2008, Shi2009, Fayolle2011b, Yokochi2012, Bar-Nun2013, Fresneau2014}. It has also been suggested that irradiation of H$_{2}$O ice may collapse pores in the amorphous structure, thereby further enhancing trapping of volatiles (e.g., \citealt{Shi2009}). Such experimental findings imply that weakly bound volatiles can be preserved in the modelled discs at temperatures that are warmer than their respective thermal desorption regimes. However, this is expected to not exceed the level of a few per cent; thus the qualitative conclusions will remain unaltered. Additional observations of interstellar ices are necessary to verify the layering of ices and to quantify the degrees of mixing and segregation, which could imply a change of the trapping medium from H$_{2}$O ice to CO$_{2}$ or CH$_{3}$OH ices, for example.

\subsubsection{Comparison with a three-phase chemical model}
To further ensure that the production of a CO$_{2}$ ice-rich zone is not a feature of the two-phase chemical model, additional modelling has been carried out with the three-phase chemical model of \citet{Furuya2015}. In the three-phase model, the ices are partitioned between a chemically active surface and an inert bulk at the expense of CPU time. For these comparative model runs, the surface is considered to consist of the uppermost $4$ monolayers. The same binding energies and initial prestellar molecular abundances (for species common to both networks), as used in the results presented thus far in this work, are assumed. For the surface layers, $E_{\text{diff}} = 0.3 E_{\text{des}}$ is used, similarly to the assumption made in the two-phase model. Further model details can be found in \citet{Furuya2015}.

Fig.~\ref{fgr:3phase} shows the percentage contribution of volatiles to the total ice content averaged for the midplane beyond $30$~AU of the infall-dominated disc at the end of the simulation ($2.46 \times 10^{5}$~yr after the precollapse phase), as obtained with the three-phase model of \citet{Furuya2015}. This figure is analogous to the right piechart of Fig.~\ref{fgr:case7_volpie}. Model results with and without the swapping mechanism are shown, which allows the exchange of molecules between the surface layers and the inert bulk ice (according to the formalism of \citealt{Garrod2013a}). In the swapping-on model, bulk ice chemistry (two-body grain-surface reactions and photochemistry) is also included with $E_{\text{diff}} = E_{\text{des}}$. Thus, the swapping-off model corresponds to a fully inert bulk; while the swapping-on setup imposes marginal chemical reactivity in the bulk ice. Upon the inclusion of swapping, weakly bound species such as CH$_{4}$ and N$_{2}$ preferentially diffuse to the surface layers and are subsequently lost to the gas phase via thermal desorption. When swapping is excluded, the quantities are consistent between the two- and three-phase models for these weakly bound species. The contributions of CH$_{3}$OH and other minor species are also in agreement.

Swapping enhances the supply of CO molecules available for reactions with OH in the surface, thus generating more CO$_{2}$ overall via the reaction with OH, as discussed earlier. This demonstrates that the same scheme that was responsible for CO$_{2}$ ice-dominance in the framework of a two-phase model persists with a three-phase model as well, once the assumption of total bulk inactivity is broken by allowing the bulk-surface exchange and marginal bulk reactivity. This implies that this is not a feature of the two-phase model, but rather a chemical effect consistent with the physical evolution of the system. The stellar FUV irradiation in the envelope during infall towards the disc is the driving force behind this CO$_{2}$ enhancement (see the spike in Fig.~\ref{fgr:phys_paper}). Since the two-phase model assumes two chemically active monolayers of the same composition as the entire mantle, reactants which may not be available in the three-phase model are available in the two-phase model. Thus, the results tend towards each other when swapping is included. The two-phase model produces even more CO$_{2}$ (an additional $7\%$) than under the assumption of a marginally reactive bulk in the swapping-on case. Currently, there is no consensus on the chemical reactivity of the bulk ice mantle and further experimental verification is necessary.

\subsection{Trace complex organic ices}
Although simple ices dominate the ice composition, there is also a variety of complex organic ices present. Table~\ref{tbl:com} quantifies the abundance of such ices relative to methanol ice for both discs and in three radial ranges. Some complex organics may become close in abundance to methanol ice, i.e., at a level of up to a few per cent of the total ice mantle. Such trace species could potentially have prebiotic implications for the synthesis of amino acids by the Strecker mechanism (e.g., \citealt{Fresneau2014, Fresneau2015}). The right panels of Figs.~\ref{fgr:case3} and~\ref{fgr:case7} show the radial distribution of complex ices and gases in the inner $50$~AU around a protostar for the spread-dominated and infall-dominated discs, respectively, at the final timestep of the evolution. Similar to the situation with simple ices, snowlines of complex organics are also recovered in the midplanes of both discs. Most are clustered near the water and methanol snowlines, as complex organics are tightly bound species (i.e., all those shown in the right panels of Figs.~\ref{fgr:case3} and~\ref{fgr:case7} except formaldehyde (H$_{2}$CO) and ketene (CH$_{2}$CO)).

For both discs, complex organic ices are most plentiful in the outer regions ($R>30$~AU). In the infall-dominated case, complex organic ices survive closer to the protostar and as far in as their respective snowline. In the spread-dominated case, complex organic ices start decreasing around $R \sim 25$~AU (except for ethanol and acetic acid). Complex organic ice formation is facilitated by weak FUV irradiation (no more than a few orders of magnitude higher than the typical cosmic ray-induced FUV flux $\sim 10^{-7}$~erg cm$^{-2}$ s$^{-1}$, \citealt{PrasadTarafdar1983}, or $\sim 10^{4}$ photons cm$^{-2}$ s$^{-1}$), such as that found for trajectories leading into more distant disc regions. Too strong FUV irradiation ($\sim G_{0}$) can quench complexity by photodissociation of complex organic ices. At the same time, luke warm temperatures are needed to enable radical mobility on grain surfaces; however, too high temperatures will result in the loss of radicals to the gas phase via thermal desorption. The warmer temperature profile of the spread-dominated disk is the main reason for the differences in complex organic abundances between the two disks. Higher temperatures explain also the larger abundance and sole survival of species like ethanol and acetic acid in the spread-dominated disc, which are considered to only form via radical-radical associations (e.g., \citealt{Walsh2014b, Walsh2014a}). Meanwhile in the infall-dominated disc, the complex organic species which are efficiently formed via hydrogenation of solid CO and derivatives are also prevalent.

\begin{landscape}
\ctable[
 caption = {Average abundances in disc midplanes of complex organic ices relative to methanol ice ($n \left( \text{X}_{\text{ice}} \right) / n \left( \text{CH$_{3}$OH}_{\text{ice}} \right) $) for three radial ranges at the end of the simulation ($2.46 \times 10^{5}$~yr after the precollapse phase) for the spread- and infall-dominated discs. The values are given for the case of the fiducial dynamic simulation and under a static assumption (i.e., constant final disc physical parameters)$^{m}$. Additionally, in the final line, the average abundance of methanol ice relative to the number of H nuclei is given for the respective radial ranges and models.},
 label = tbl:com
 ]{@{\extracolsep{\fill}}lllllllllllll}{
 \tnote[m]{The notation $a\left(b\right)$ stands for $a \times 10^{b}$.}
 }{
 \hline
 R (AU) & \multicolumn{4}{c}{$1-10$} & \multicolumn{4}{c}{$10-30$} & \multicolumn{4}{c}{$>30$} \T\B\\
 \hline
 \multirow{2}{*}{\diagbox[height=3em,trim=l]{Species}{disc}} & \multicolumn{2}{c}{spread-dom.} & \multicolumn{2}{c}{infall-dom.} & \multicolumn{2}{c}{spread-dom.} & \multicolumn{2}{c}{infall-dom.} & \multicolumn{2}{c}{spread-dom.} & \multicolumn{2}{c}{infall-dom.} \T\\
 \cline{2-13}
   & \multicolumn{1}{c}{dyn.} & \multicolumn{1}{c}{stat.} & \multicolumn{1}{c}{dyn.} & \multicolumn{1}{c}{stat.} & \multicolumn{1}{c}{dyn.} & \multicolumn{1}{c}{stat.}& \multicolumn{1}{c}{dyn.} & \multicolumn{1}{c}{stat.} & \multicolumn{1}{c}{dyn.} & \multicolumn{1}{c}{stat.} & \multicolumn{1}{c}{dyn.} & \multicolumn{1}{c}{stat.} \T\B\\ 
 \hline
 H$_{2}$CO         & $4.7 \left( -7 \right)$ & $3.2 \left( -7 \right)$ & $3.4 \left( -4 \right)$ & $4.6 \left( -6 \right)$
                   & $1.4 \left( -2 \right)$ & $1.7 \left( -3 \right)$ & $7.1 \left( -5 \right)$ & $3.3 \left( -1 \right)$
                   & $1.4 \left( -2 \right)$ & $1.5 \left( -2 \right)$ & $2.4 \left( -1 \right)$ & $1.1 \left( -3 \right)$ \T \\
 CH$_{2}$CO        & $1.7 \left( -6 \right)$ & $1.1 \left( -6 \right)$ & $9.8 \left( -5 \right)$ & $2.1 \left( -5 \right)$
                   & $5.7 \left( -3 \right)$ & $1.4 \left( -2 \right)$ & $4.6 \left( -3 \right)$ & $3.9 \left( -2 \right)$
                   & $2.9 \left( -2 \right)$ & $7.7 \left( -2 \right)$ & $1.3 \left( -2 \right)$ & $2.0 \left( -2 \right)$ \\
 HCOOH             & $1.6 \left( -7 \right)$ & $3.9 \left( -7 \right)$ & $1.7 \left( -5 \right)$ & $2.2 \left( -7 \right)$
                   & $1.1 \left( -3 \right)$ & $3.7 \left( -5 \right)$ & $2.6 \left( -3 \right)$ & $3.0 \left( -3 \right)$
                   & $2.0 \left( -4 \right)$ & $4.8 \left( -3 \right)$ & $4.1 \left( -1 \right)$ & $7.0 \left( -4 \right)$ \\
 HCOOCH$_{3}$      & $2.7 \left( -4 \right)$ & $1.6 \left( -5 \right)$ & $2.4 \left( -2 \right)$ & $3.7 \left( -5 \right)$
                   & $3.9 \left( -2 \right)$ & $2.5 \left( -4 \right)$ & $5.1 \left( -3 \right)$ & $1.9 \left( -2 \right)$
                   & $1.2 \left( -2 \right)$ & $8.3 \left( -4 \right)$ & $2.4 \left( -2 \right)$ & $3.0 \left( -3 \right)$ \\
 CH$_{3}$CHO       & $1.7 \left( -9 \right)$ & $2.7 \left( -9 \right)$ & $1.7 \left( -8 \right)$ & $1.6 \left( -7 \right)$
                   & $1.6 \left( -4 \right)$ & $1.5 \left( -6 \right)$ & $1.1 \left( -3 \right)$ & $1.9 \left( -1 \right)$
                   & $1.3 \left( -2 \right)$ & $1.6 \left( -3 \right)$ & $7.7 \left( -3 \right)$ & $8.7 \left( -4 \right)$ \\
 CH$_{3}$OCH$_{3}$ & $1.4 \left( -6 \right)$ & $1.4 \left( -2 \right)$ & $1.6 \left( -3 \right)$ & $1.0 \left( -2 \right)$
                   & $1.1 \left( -3 \right)$ & $2.6 \left( -2 \right)$ & $6.8 \left( -3 \right)$ & $2.9 \left( -2 \right)$
                   & $1.5 \left( -2 \right)$ & $7.1 \left( -3 \right)$ & $1.5 \left( -2 \right)$ & $1.4 \left( -2 \right)$ \\
 C$_{2}$H$_{5}$OH  & $3.7 \left( -1 \right)$ & $1.8 \left( -2 \right)$ & $2.7 \left( -1 \right)$ & $1.8 \left( -2 \right)$
                   & $2.0 \left( -1 \right)$ & $3.1 \left( -2 \right)$ & $3.8 \left( -2 \right)$ & $2.2 \left( -2 \right)$
                   & $1.7 \left( -2 \right)$ & $2.2 \left( -2 \right)$ & $2.6 \left( -2 \right)$ & $1.3 \left( -2 \right)$ \\
 CH$_{3}$COOH      & $9.8 \left( -1 \right)$ & $3.3 \left( -1 \right)$ & $4.4 \left( -1 \right)$ & $3.5 \left( -1 \right)$
                   & $2.1 \left( -1 \right)$ & $2.0 \left( -1 \right)$ & $1.7 \left( -2 \right)$ & $2.6 \left( -2 \right)$
                   & $5.7 \left( -3 \right)$ & $2.4 \left( -3 \right)$ & $4.4 \left( -1 \right)$ & $2.1 \left( -3 \right)$ \\
 HOCH$_{2}$CHO     & $3.2 \left( -4 \right)$ & $1.0 \left( -3 \right)$ & $1.5 \left( -1 \right)$ & $1.1 \left( -2 \right)$
                   & $1.0 \left(~~~0\right)$ & $2.3 \left( -1 \right)$ & $2.3 \left( -1 \right)$ & $8.6 \left( -1 \right)$
                   & $4.2 \left( -1 \right)$ & $4.6 \left( -1 \right)$ & $2.1 \left( -1 \right)$ & $5.8 \left( -2 \right)$ \B\\
 \hline
 $n \left( \text{CH$_{3}$OH}_{\text{ice}} \right) / n_{\text{H}}$
                   & $2.8 \left( -6 \right)$ & $1.9 \left( -6 \right)$ & $5.1 \left( -6 \right)$ & $2.0 \left( -6 \right)$
                   & $5.2 \left( -6 \right)$ & $2.5 \left( -6 \right)$ & $8.5 \left( -6 \right)$ & $3.0 \left( -6 \right)$
                   & $5.4 \left( -6 \right)$ & $5.8 \left( -6 \right)$ & $1.7 \left( -6 \right)$ & $3.0 \left( -5 \right)$ \T\B\\
 \hline
 }
\end{landscape}

For both discs an inner reservoir rich in gas-phase complex organics is recovered in these models. This zone appears to be the inner few AU of the disc midplanes, where the temperatures are high enough to thermally desorb all tightly bound species; and shielding by dust is sufficiently strong to preserve these molecules from the protostar's photodissociating FUV photons.

The abundances in Table~\ref{tbl:com} show that complex organic ices are plentiful in protoplanetary disks, especially in the outer regions. The current model is, however, tailored to the most favourable conditions for complex organic molecule production. This is achieved by assuming $E_{\text{diff}} = 0.3 E_{\text{des}}$ \citep{HasegawaHerbstLeung1992}, which allows efficient radical-radical chemistry on grain surfaces. A two-phase model, additionally, assumes the same composition for the two chemically active monolayers as for the entire mantle, thus increasing the availability of species for radical-radical chemistry without accounting for, as an example, lower radical mobility deeper in the mantle. Hence, the presented abundances may be seen as upper limits on the quantities of complex organic ices in Class $0$/I protoplanetary disk midplanes. These aspects are explored in a detailed parameter study in \citet{Drozdovskaya2015}. On the other hand, the abundances of complex organic molecules may be enhanced by additional low ($\sim 15$~K) temperature formation routes that are seen in recent laboratory studies even in the absence of FUV radiation (e.g., \citealt{Fedoseev2015, Chuang2016}).

The predicted ice reservoirs can potentially be traced indirectly via cometary studies, which are showing that comets are rich in large complex organic species. For example, the highly complex molecule glycine is detected at a ratio of $0.5$ relative to methanol on 67P \citep{Altwegg2016, LeRoy2015}. This implies that the modelled abundances are not very different from such observational evidence. A more detailed comparison with comets is presented in Section~\ref{comets}. Observations of hot inner regions (so-called hot cores or corinos) place the gas-phase abundances of methyl formate, dimethyl ether, ethanol and glycolaldehyde at $\sim 1-10^{-2}, 1-10^{-2}, 10^{-1}-10^{-2}, 10^{-2}-10^{-3}$ relative to gaseous methanol, respectively (fig.~$8$ of \citealt{Taquet2015}). These values are comparable to those in Table~\ref{tbl:com} with the exception of glycolaldehyde, which is most likely caused by the lack of destructive chemical pathways in the network, while already including additional formation pathways via glyoxal \citep{Drozdovskaya2015}. Comparisons of the modelled complex organic ice abundances in disk midplanes with gaseous abundances in hot cores/corinos should, however, be made with caution since it remains unclear what the origin of the hot core emission is. In hot cores, thermal desorption leaves desorbed molecules intact, but the observed abundances are potentially modified by gas-phase reactions, since the temperatures and densities are high. Typically, ratios of isomers vary between hot core observations and models, which could indicate either missing chemical pathways in the network or the importance of gas-phase reactions in hot cores. Meanwhile, observed gases in (outer) disk regions are released via non-thermal desorption processes, which may break molecules apart \citep{CruzDiaz2016, Bertin2016, Walsh2016}.

\subsection{Comparison to planet population synthesis models}
\citet{Marboeuf2014b} studied the volatiles of planetesimals in a static cooling protoplanetary disc. The cooling occurs, because the disc surface density decreases with time, as a consequence of material accreting on to the star. As explained in Section~\ref{coldhot}, the models of \citet{Marboeuf2014b} assume a hybrid of the `hot' and `cold' start scenarios. The volatiles are assumed to consist of H$_{2}$O, CO, CO$_{2}$, CH$_{3}$OH, CH$_{4}$, N$_{2}$, NH$_{3}$ and H$_{2}$S and are the sole potential ice constituents (if disc conditions allow the solid phase), because they are observed to be dominant interstellar and cometary volatiles. All other species are ignored, which means that as much as $\sim 10$\% of the total ice is possibly missed (Figs.~\ref{fgr:case3_volpie} and~\ref{fgr:case7_volpie}).

Taking these assumptions about volatiles, the hybrid start scenario, the use of phase diagrams and the differences in disc models into account, there are still some similarities in the findings of \citet{Marboeuf2014b} and the dynamic models presented in this work. Here, the inner discs in the embedded phase are somewhat more massive than those of \citet{Marboeuf2014b}, where $\Sigma \lesssim 500$~g~cm$^{-2}$ at $R=5.2$~AU, while here $\Sigma \sim 630$~g~cm$^{-2}$ and  $\Sigma \sim 540$~g~cm$^{-2}$ for the spread-dominated and infall-dominated discs at $R=5.2$~AU, respectively ($\Sigma \sim 110$~g~cm$^{-2}$ and  $\Sigma \sim 300$~g~cm$^{-2}$ for the spread-dominated and infall-dominated discs at $R=30$~AU; Fig.~\ref{fgr:midplanes}). First, similar trends in snowlines are observed. Most volatiles thermally desorb in the inner few AU ($\lesssim 10$~AU). The most volatile species, CH$_{4}$, CO and N$_{2}$, freeze out further out in the midplanes, if at all, such as in the hotter irradiated disc model of \citet{Marboeuf2014b} and the spread-dominated disc of this work. This consequently leads to the agreement in water being the dominant ice constituent in planetesimals at $\sim 55-75$\% of the total ice content, with its contribution increasing with decreasing radius, as molecules that are more weakly bound thermally desorb in accordance with the midplane temperature profile (table~$9$ and figs.~$15$,~$16$ of \citealt{Marboeuf2014b}). The denominator used to obtain the values given in \citet{Marboeuf2014b} accounts only for $8$ molecules, while in this work all available solids are summed to obtain the denominator. Further quantitative comparison of the planetesimal ice composition, as obtained via full prestellar inheritance by discs and subsequent disc cooling \citep{Marboeuf2014b}, versus that via full kinetic chemical modelling and dynamic collapse (this work) is not possible within the framework of our models.

Instead, the assumption of complete prestellar inheritance by discs can be tested by contrasting the assumed initial volatile abundances of \citet{Marboeuf2014b} with the midplane abundances obtained in this work. Table~\ref{tbl:Marboeuf_compar} shows the modelled midplane ice composition for three sets of initial abundances: the model-consistent fiducial prestellar phase abundances; and the CO-poor (CO:CO$_{2}=1:1$) and CO-rich (CO:CO$_{2}=5:1$) sets, which are based on observations of prestellar cores and protostellar sources as justified in \citet{Marboeuf2014b}. Table~\ref{tbl:Marboeuf_compar} contains the results for both the spread- and infall-dominated disc midplanes. All values are number averages and focus on the most ice-rich outer regions of the midplanes beyond $30$~AU. By comparing the tabulated initial and final disc values, it can be seen that the largest differences lie in the amount of CO$_{2}$ produced. In this work, more CO$_{2}$ is produced via the reaction of OH and CO mentioned earlier, which in turn eats away at the water and CO ice reservoirs. For the infall-dominated case with an excess of CO (CO:CO$_{2}=5:1$), this effect is at its most extreme as water ice is almost entirely converted to CO$_{2}$ ice. Another signficant change is seen for CH$_{3}$OH ice. If more than the fiducial prestellar quantity is injected into the system initially, as under the CO-poor and CO-rich conditions, then the final disc contribution is still $\lesssim 3$ per cent, as with the fiducial input. Methanol ice is chemical processed en route to the disc via photon-induced chemistry and used to build larger, more complex molecules, leading to a disc abundance of a few per cent that is independent of the initial abundance.

\subsubsection{Dependence on initial conditions}

\ctable[
 star = 1,
 caption = Icy volatiles in outer ($R > 30$~AU) disc midplanes for different initial abundances at the end of the simulation ($2.46 \times 10^{5}$~yr after the precollapse phase) for the spread- and infall-dominated discs for models including dynamic collapse and those for which static conditions were assumed.,
 label = tbl:Marboeuf_compar
 ]{@{\extracolsep{\fill}}llllrrrrr}{
 \tnote[n]{Under the precollapse phase scenario, initial abundances are split over the gas and solid phases. Under the two sets of initial conditions adopted from \citet{Marboeuf2014b}, the abundances are set for the gas-phase only in line with the `hot' disc assumption.}
 \tnote[o]{Number averages for $R > 30$~AU are given, as that region is the richest in ices. Note that the radial range considered for the infall-dominated disc is larger since the average is computed from $R=30$~AU and out to the outer radius, which is much larger than for the spread-dominated disc.}
 \tnote[p]{If all the gases deposit as ices.}
 \tnote[q]{A few per cent can be trapped in H$_{2}$O ice (Section~\ref{binding}).}
 }{
 \hline
 \multirow{2}{*}{Initial assumption} & \multirow{2}{*}{Species} & \multicolumn{2}{c}{Initial abundance$^{n}$} & \multicolumn{5}{c}{$n \left( \text{X}_{\text{ice}} \right) / \sum n \left( \text{X}_{\text{ice}} \right)$ (\%)$^{o}$} \T\B\\
 \cline{3-9}
   &  & \multirow{2}{*}{$n \left( \text{X}_{\text{gas}} \right) / n_{\text{H}}$} & \multirow{2}{*}{$n \left( \text{X}_{\text{ice}} \right) / n_{\text{H}}$} & \multirow{2}{*}{initial} & \multicolumn{2}{c}{spread-dom.} & \multicolumn{2}{c}{infall-dom.} \T\\
 \cline{6-9}
   &  &  &  &  & dyn. & stat. & dyn. & stat. \T\B\\
 \hline
 \multirow{8}{*}{Prestellar}      & H$_{2}$O   & $7.7\times10^{-8}$  & $1.9\times10^{-4}$ & $51~$    & $69~$   & $57~$   & $35~$   & $51~$\T\\
                                  & CO         & $1.4\times10^{-5}$  & $5.9\times10^{-5}$ & $16~$    & $0^{q}$ & $0^{q}$ & $0^{q}$ & $1~$\\
                                  & CO$_{2}$   & $7.9\times10^{-8}$  & $1.9\times10^{-5}$ & $5~$     & $14~$   & $27~$   & $46~$   & $21~$\\
                                  & NH$_{3}$   & $2.1\times10^{-7}$  & $4.8\times10^{-6}$ & $1~$     & $4~$    & $3~$    & $1~$    & $4~$\\
                                  & CH$_{4}$   & $1.2\times10^{-7}$  & $1.5\times10^{-5}$ & $4~$     & $0^{q}$ & $0^{q}$ & $4~$    & $3~$\\
                                  & CH$_{3}$OH & $2.0\times10^{-10}$ & $3.7\times10^{-6}$ & $1~$     & $2~$    & $2~$    & $1~$    & $10~$\\
                                  & H$_{2}$S   & $5.3\times10^{-10}$ & $8.1\times10^{-9}$ & $0^{q}$  & $0^{q}$ & $0^{q}$ & $0^{q}$ & $0^{q}$\\
                                  & N$_{2}$    & $1.3\times10^{-5}$  & $1.1\times10^{-5}$ & $3~$     & $0^{q}$ & $0^{q}$ & $3~$    & $2~$\B\\
 \hline
 \multirow{8}{*}{CO:CO$_{2}=1:1$} & H$_{2}$O   & $1.2\times10^{-3}$  & -                  & $57^{p}$ & $63~$   & -       & $27~$   & -\T\\
                                  & CO         & $2.4\times10^{-4}$  & -                  & $11^{p}$ & $0^{q}$ & -       & $0^{q}$ & -\\
                                  & CO$_{2}$   & $2.4\times10^{-4}$  & -                  & $11^{p}$ & $26~$   & -       & $48~$   & -\\
                                  & NH$_{3}$   & $8.3\times10^{-5}$  & -                  & $4^{p}$  & $3~$    & -       & $0^{q}$ & -\\
                                  & CH$_{4}$   & $7.1\times10^{-5}$  & -                  & $3^{p}$  & $0^{q}$ & -       & $4~$    & -\\
                                  & CH$_{3}$OH & $1.8\times10^{-4}$  & -                  & $8^{p}$  & $3~$    & -       & $0^{q}$ & -\\
                                  & H$_{2}$S   & $2.4\times10^{-5}$  & -                  & $1^{p}$  & $0^{q}$ & -       & $0^{q}$ & -\\
                                  & N$_{2}$    & $8.3\times10^{-5}$  & -                  & $4^{p}$  & $0^{q}$ & -       & $5~$    & -\B\\
 \hline
 \multirow{8}{*}{CO:CO$_{2}=5:1$} & H$_{2}$O   & $8.3\times10^{-4}$  & -                  & $39^{p}$ & $62~$   & -       & $1~$    & -\T\\
                                  & CO         & $8.2\times10^{-4}$  & -                  & $39^{p}$ & $0^{q}$ & -       & $16~$   & -\\
                                  & CO$_{2}$   & $1.6\times10^{-4}$  & -                  & $8^{p}$  & $27~$   & -       & $73~$   & -\\
                                  & NH$_{3}$   & $5.7\times10^{-5}$  & -                  & $3^{p}$  & $3~$    & -       & $0^{q}$ & -\\
                                  & CH$_{4}$   & $4.9\times10^{-5}$  & -                  & $2^{p}$  & $0^{q}$ & -       & $0^{q}$ & -\\
                                  & CH$_{3}$OH & $1.2\times10^{-4}$  & -                  & $6^{p}$  & $3~$    & -       & $0^{q}$ & -\\
                                  & H$_{2}$S   & $1.6\times10^{-5}$  & -                  & $1^{p}$  & $1~$    & -       & $0^{q}$ & -\\
                                  & N$_{2}$    & $5.7\times10^{-5}$  & -                  & $3^{p}$  & $0^{q}$ & -       & $3~$    & -\B\\
 \hline}
 
Table~\ref{tbl:Marboeuf_compar} also gives insight into how the modelled disc abundances depend on the initial chemical conditions used. If all the gaseous volatiles in \citet{Marboeuf2014b} are assumed to deposit as ices, then, overall, more volatiles are frozen out than when a model-consistent prestellar phase is calculated. For the CO-poor abundances, the initial volatile quantities are increased by factors of $6.3$, $4.3$, $13$, $17$, $4.7$, $49$, $3.0 \times 10^{3}$, $7.5$ for H$_{2}$O, CO, CO$_{2}$, NH$_{3}$, CH$_{4}$, CH$_{3}$OH, H$_{2}$S and N$_{2}$, respectively, relative to the fiducial presteller ice abundances. The largest increases, thus, are for CH$_{3}$OH and H$_{2}$S. The CO-rich abundances are extreme with CO ice being available at the level of H$_{2}$O ice.

For the spread-dominated disc, it is then seen that if more CO$_{2}$ is present initially with the CO-poor abundances, then that quantity is preserved and more CO$_{2}$ is made en route as also seen with the model-consistent prestellar phase abundances. If more CO is provided initially with the CO-rich abundances, then it is efficiently converted to CO$_{2}$ via the reaction with OH. In both situations, this leads to CO$_{2}$ occupying a more significant portion of the ice mantle than under those of the prestellar phase. This also holds for the infall-dominated disc. Since CO$_{2}$ was already abundant in the outer regions of that disc, under the CO-poor and CO-rich initial conditions favoring larger quantities of CO$_{2}$, the contribution of CO$_{2}$ to the ice mantle is even more extreme.

\subsubsection{Static scenario}
\label{static}
To further test the importance of dynamic motions in forming protostellar systems, a static scenario has also been computed. Material is assumed to be at the constant final physical conditions of the midplane for the entire duration of the evolution, i.e., for $2.46 \times 10^{5}$~yr, as is tradionally assumed for Class II disc models. For both discs, the largest impact is seen for complex organics (Table~\ref{tbl:com} for $R > 30$~AU). Fewer of the largest complex organic molecules are produced in both discs under a static assumption. By taking dynamic infall into account, protoplanetary disc materials are exposed to elevated temperature and increased FUV fluxes en route to the disc, which facilitates radical production and mobility on the grain surfaces, both of which are key to forming larger species in these models. This is clearly illustrated in Fig.~\ref{fgr:phys_paper}, since the abundances of complex organics increase along with the FUV flux increase in the envelope. Under the static assumption for the infall-dominated disc in the $10-30$~AU range, the most favorable conditions for complex organic formation appear to be shifted from the outer disc to a more inner region. This in turn affects the availability of methanol ice, which is underproduced by a factor of $10$ for this zone in comparison to the dynamic model runs (not shown).

The static assumption also shifts the balance of H$_{2}$O and CO$_{2}$ ices in the outer regions of the two discs (Table~\ref{tbl:Marboeuf_compar}). For the spread-dominated disc, the static scenario means cooler temperatures for the duration of the simulation, leading to a higher availability of CO on the grains (if only transiently), which leads to more CO$_{2}$ ice at the expense of H$_{2}$O ice. For the infall-dominated disc, the static scenario means lower FUV irradiation, leading to less frequent H$_{2}$O ice photodissociation and thus a reduction of OH, on the contrary reducing the amount of CO$_{2}$ ice contribution in the outer disc. Other volatiles show variations of no more than a few per cent between the static and dynamic cases.

\section{Discussion}
\label{discussion}
\subsection{Dynamics, chemistry and inheritance}
In this work, the importance of dynamic collapse and kinetic chemistry en route to protoplanetary disc midplanes has been assessed by considering two physical models and by running various test cases. The two modelled discs predominantly grow by different mechanisms, either via viscous spreading or by pure infall, leading to material being transported along different trajectories. The results have shown that there is a delicate balance between H$_{2}$O and CO$_{2}$ ices. Under cool ($\sim 20$~K) temperatures, allowing the presence of CO ice, and weak FUV irradiation, producing the OH radical from the photodissociation of H$_{2}$O ice, a significant water-carbon dioxide imbalance can occur leading to large quantities of CO$_{2}$ ice at the expense of H$_{2}$O (and CO) ice(s). Such conditions are encountered at large radii ($R>30$~AU) for the infall-dominated disc ($\sim 46$\% of the ice mantle is CO$_{2}$). Under an assumption of a static disc, this also occurs for the outer region of the spread-dominated disc, while the outer region of the infall-dominated disc becomes less imbalanced ($\sim 25$\% of the ice mantle is CO$_{2}$). If either the chemistry or the dynamic transport of volatiles was neglected, then these shifts in the ice mantle composition would be missed. This effect has been shown to be insensitive to chemical model parameters such as the diffusion and quantum tunnelling barriers, and binding energies. It also manifests when three-phase models are used. If the FUV field during the physical evolution would be modified, then the qualitative results would still hold; and the amount of CO$_{2}$ ice produced would only change quantitatively. In the case of a weaker FUV field, the amount of CO$_{2}$ produced would decrease (by at most $\sim 25$ and $\sim 13$ per cent for the infall- and spread-dominated cases, based on the static case without any FUV exposure in the envelope). In the case of a stronger FUV field, the amount of CO$_{2}$ is expected to increase, since more FUV photons are available to dissociate H$_{2}$O ice. The FUV field can also be harder or softer, and photodissociation cross-sections are wavelength-dependent, thus potentially shifting regional enhancements. Given the importance of the FUV flux demonstrated here, this should be explored in future work.

The trace complex organic species also show a strong dependence on dynamics. Their production is a delicate process balanced by dust temperatures that are warm enough to allow radical mobility on grain surfaces, but low enough to make radical thermal desorption inefficient; and by FUV irradiation that is sufficiently strong to produce radicals in the ice, but sufficiently weak to not efficiently photodissociate the complex organics back into radicals. As a result, dynamic infall leading to elevated temperatures and increased FUV exposure, facilitates complex organic molecule production, which would otherwise be impeded by the midplane physical conditions. This also influences the availability of methanol ice, which is a key parent species for many complex organics, making it sensitive to dynamic transport and the chemistry en route. The outer zones ($R>30$~AU) of both the spread- and infall-dominated discs contain the most complex organic ices (at abundances as high as that of methanol ice, i.e., at a level of up to a few per cent of the total ice mantle), while the inner discs (inner few AU) are rich in gaseous complex organics. These molecules can only be modelled by using full kinetic chemical calculations.

Other volatiles are found to be insensitive to the dynamics and chemistry between the cloud and disc phases. The findings made with these physicochemical models are summarized in Fig.~\ref{fgr:scenarios}. The implication is that not all prestellar ices are simply inherited by the midplanes of protoplanetary discs. The balance between dominating volatiles may change depending on the route taken by the material to get to the planetesimal- and comet-forming zones and the chemistry that occurs during that time. The variations seen in H$_{2}$O, CO$_{2}$ and CH$_{3}$OH ices are consistently obtained under different initial volatile abundances. The complex organic molecule production is facilitated by the favorable conditions in the envelope, thus determining the prebiotically-significant composition of planetesimals outside of the protoplanetary disc.

\subsection{Comparison to other disc models}
A large number of models exist for the Class II protoplanetary discs of varying physical and chemical complexity. Therefore, in this section, the models of young embedded discs will only be compared to the more evolved discs from studies including chemistry of complex organic molecules. The series of papers \citet{Walsh2014b, Walsh2014a}, \citet{Walsh2015} have studied a full range of molecules in protoplanetary discs under static conditions, along accretion flows within the disc, and for various central irradiating protostars, using the same prescription for the chemistry as that used in this work.

Comparing with the midplane abundances of complex organics for the static Class II disc of \citet{Walsh2014a}, it appears that the spead-dominated disc contains less acetaldehyde ($\sim 2$ orders of magnitude) and more confined to the outer region ($R \gtrsim 30$~AU) rather than up to the thermal desorption front; less dimethyl ether ($\sim 1$ order of magnitude), but more acetic acid ($\sim 2$ orders of magnitude). The infall-dominated disc seems to be richer in complex organic molecules than the static Class II disc (more formic acid, methyl formate and acetic acid, all by $\sim 1$ order of magnitude). These chemical variations stem from different physical conditions and time-scales of static Class II discs in comparison to these dynamically formed embedded Class I discs. Comparing with the $z=0$~AU accretion flow through the disc of \citet{Walsh2014b}, it seems that the spead-dominated disc abundances lie somewhere inbetween the isolated and extremely irradiated disc models (formic acid abundances are closer to the extremely irradiated disc, while methyl formate abundances are closer to the isolated disc, and acetaldehyde abundances lie inbetween the two models). The infall-dominated disc abundances are very similar to those in the midplane accretion flow of the isolated disc with only variations of factors of a few. Due to the inclusion of the additional pathway to glycolaldehyde via glyoxal in this work, more glycolaldehyde is made here by $\sim 2$ orders of magnitude.

Comparing the spread- and infall-dominated discs to the results from turbulent disc models of \citet{Furuya2014}, it appears that formaldehyde, methanol and ammonia agree within a factor of a few with the non-turbulent ($\alpha_{z}=0$) model. However, there are subtle differences for complex organics such as methyl formate and dimethyl ether. Here, the outer disc in its entirety is rich in large complex species, however in the models of \citet{Furuya2014} the distribution of such species is highly localized. The abundances peak and match those obtained in this work only in a narrow radial range. This is most likely because \citet{Furuya2014} assume $E_{\text{diff}} = 0.5 E_{\text{des}}$, which inhibits efficient complex organic molecule formation (see appendix~A of \citealt{Walsh2014a}). These comparisons imply that current Class II disc models may be underestimating the complex organic content, since the initial chemical conditions set by the preceeding embedded phase could be much higher than currently assumed.

\subsection{Implications for population synthesis models}

\begin{figure*}
 \centering
 \begin{subfigure}[b]{0.45\textwidth}
 \includegraphics[width=\textwidth,keepaspectratio]{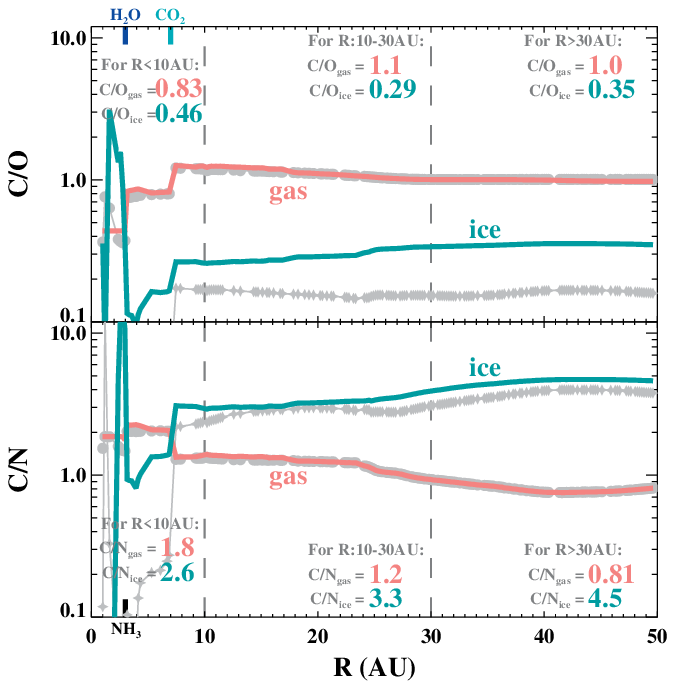}
 \caption{Spread-dominated disc}
 \label{fgr:case3COCN}
 \end{subfigure}
 \begin{subfigure}[b]{0.45\textwidth}
 \includegraphics[width=\textwidth,keepaspectratio]{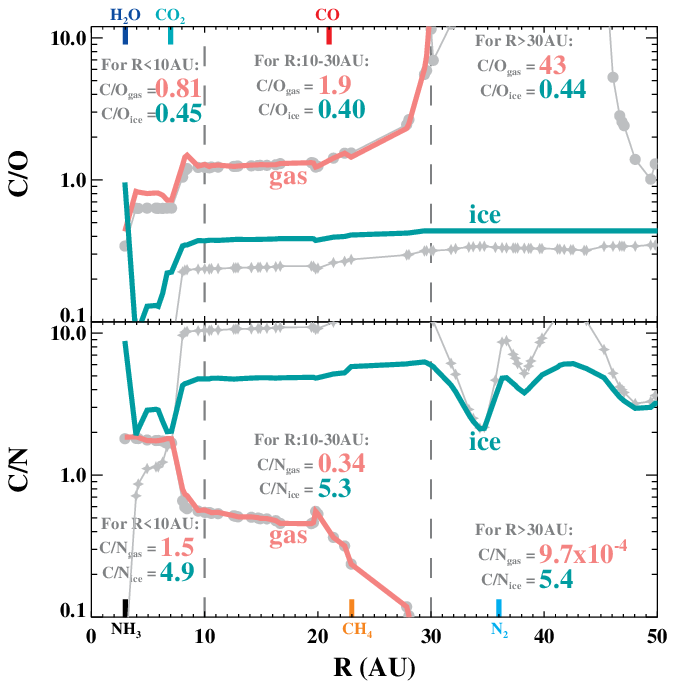}
 \caption{Infall-dominated disc}
 \label{fgr:case7COCN}
 \end{subfigure}
 \caption{The midplane C/O and C/N ratios as a function of disc radius ($R$) in the gaseous (coral colored lines) and solid (torquoise colored lines) phases at the end of the simulation ($2.46 \times 10^{5}$~yr after the onset of collapse). The average ratios over three radial ranges (separated by grey dashed lines) are provided; and the locations of some snowlines are marked (the full list is in Table~\ref{tbl:snowlines}). The grey curves with dots and stars are the ratios in the gasesous and solid phases, respectively, as calculated with abundances of simple volatiles only (namely H$_{2}$O, CO, CO$_{2}$, CH$_{3}$OH, CH$_{4}$, N$_{2}$ and NH$_{3}$.)}
\end{figure*}

The results of these models on midplane volatiles have been compared to the initial volatile composition of planetesimals used in sophisticated planet population synthesis models of \citet{Marboeuf2014b, Marboeuf2014a}, \citet{Thiabaud2014, Thiabaud2015}. The two sets of models and assumptions are put in perspective in Fig.~\ref{fgr:scenarios}. The main point is that complete inheritance of cloud volatiles by discs is not a valid assumption for species such as H$_{2}$O, CO$_{2}$ and CH$_{3}$OH for all regions of the disc, which have been shown to be affected by the dynamics and chemistry during transport. Thus, under current assumptions, planet population synthesis models are potentially overestimating the availability of CH$_{3}$OH by a factor of $\sim 2-4$ and could be missing large regional CO$_{2}$ reservoirs. Complete exclusion of complex organic ices means that as much as $\sim 10$\% of the total ice may be currently omitted.

If the results of these models were to be used as the initial volatile abundances for population synthesis models, then the initial C/O and C/N ratios of planetesimals would be altered (the ratios denote the overall volatile carbon, oxygen, and nitrogen budgets). The midplane gaseous and solid C/O and C/N ratios are shown as a function of disc radius for the spread-dominated disc in Fig.~\ref{fgr:case3COCN} and for the infall-dominated disc in Fig.~\ref{fgr:case7COCN}. The first major point is that the two ratios in ices do not match those in gases, i.e., the two phases are decoupled in terms of these tracers. These findings are fairly consistent with those in fig.~$1$ of \citet{Oberg2011COratio}, which were based on a much simpler adsorption-thermal desorption prescription for a static disc temperature profile, with the exception of the large gas-phase C/O jump seen in the outer regions of the infall-dominated disc.

For the spread-dominated disc, the gas-phase C/O ratio $\sim 1$. The ratio decreases to $\sim 0.8$ only in the inner $10$~AU, once CO$_{2}$ and H$_{2}$O ices sequentially thermally desorb and increase the oxygen budget in the gas. The solid-phase C/O ratio in this warmer ($\sim 40$~K) disc is subsolar at a value $\sim 0.3$, which reflects the ices being dominated by water and carbon dioxide setting C/O $\sim 1/3$. The ratio reaches towards the solar value of $0.5$ in the inner few AU once CO$_{2}$ and H$_{2}$O are lost as gases. For the infall-dominated disc, the gaseous ratio is comparable to that of the spread-dominated disc for the inner $30$~AU (with a value in the $\sim 0.8-2$ range). However, beyond that radius, the ratio increases by an order of magnitude as a result of CO adsorbing and locking away almost all oxygen carriers. The solid-phase ratio nearly matches the solar C/O ratio with values in the $0.4-0.45$ range along the entire extent of the disc, which stems from the larger CO$_{2}$ reservoir seen for in this disc, which drives the ratio towards $1/2$.

The C/N ratio in the spread-dominated disc varies roughly between $0.8$ and $1.8$ in the gas and between $2.5$ and $4.5$ in the ice, thus below the solar value of $4.9$ in both phases. The biggest variations are primarily driven by the thermal desorption events of the main icy carbon carrier CO$_{2}$ and the main icy nitrogen carrier NH$_{3}$ under the warm $\sim 40$~K conditions of this midplane. Beyond $30$~AU, where complex organic ices are plentiful, the fraction of carbon stored in the solid phases increases and so does the C/N ratio in the ice. In the infall-dominated disc, the C/N ratio in the ice is approximately solar for all radii. The gas-phase ratio rapidly decreases beyond $20$~AU, since the majority of both C and N carriers are frozen out at the cold $\sim 20$~K conditions of this disc. The solid phase C/N ratio in this disc is in agreement with the observed cometary value of $\sim 7$, however it does not explain the ratios on the order of $\sim 50$ for the Earth and chondrites in the inner few AU (see fig.~$1$ of \citealt{Bergin2015}).

In Figs.~\ref{fgr:case3COCN} and~\ref{fgr:case7COCN}, gray lines depict the C/O and C/N ratios as calculated based on the abundances of the main volatiles only, in contrast to the colored lines, which are calculated based on the full set of species in the network and model-consistent prestellar initial abundances. For the gas phase, the differences between the two calculations are minor (except the very inner $1-2$~AU). On the other hand, for the solid phase, variations by a factor of $2-3$ are seen. These differences reflect the quantities of C, O and N locked away in complex organic ices and demonstrate the importance of the $\sim 10$ per cent of the icy mantle labelled as `other' in the earlier discussions. The implications of the volatile budget derived with these models for the planet populations are the subject of future collaborative work, which will account for drift beyond the traditional inner $30$~AU planet-forming zone. Such work will show if chemical processing during the earlier dynamic phases of protoplanetary disc formation may help explain CO$_{2}$-rich exoplanet atmospheres, which cannot be replicated with chemical equilibrium models at atmospheric temperatures and pressures \citep{HengLyons2016}, although chemical processing within the atmospheres themselves is non-negligible. The differences in the C/O and C/N ratios between the two phases could potentially be used to trace the contribution of ices from planetesimals to planetary atmospheres versus that of ambient disc gas.

\subsection{Comparison with comets}
\label{comets}
Comets are thought to spend the dominant portion of their lifetime at large radii, far away from the heat and irradiation from the Sun. As a result, they are considered to be the most pristine probes of the composition of the young protosolar disc. Observed chemical similarities between comets and protostellar environments are puzzling (see \citealt{MummaCharnley2011} and \citealt{CaselliCeccarelli2012} for reviews). It remains unclear whether the agreements signify comets inheriting materials from earlier phases of star and planet formation, or if it simply is a general coincidence caused by chemistry proceeding along similar pathways under astrophysical conditions. The modelled midplane volatile contributions relative to water ice generally agree with the observed ranges for cometary volatiles within a factor of $2$ (comparing the values quoted in Figs.~\ref{fgr:case3_volpie} and~\ref{fgr:case7_volpie} with those in fig.~$4$ of \citealt{MummaCharnley2011}). The dominating ice constituents are found to be H$_{2}$O, CO$_{2}$, NH$_{3}$, CH$_{4}$ and CH$_{3}$OH consistently. Only CO is underproduced in our models, but as mentioned earlier, this could be resolved by trapping in amorphous water ice (Section~\ref{binding}). The CO$_{2}$ ice-rich zone seen in this work can help explain CO$_{2}$-rich comets (fig.~$4$ of \citealt{MummaCharnley2011}), and the CO$_{2}$-rich and H$_{2}$O-poor jets of comet 103P/Hartley 2 \citep{AHearn2011}.

Comets are rich in many complex organics; and the recent measurements for 67P/Churyumov-–Gerasimenko with the \textit{Rosetta} mission are revealing even more complexity than suggested by earlier observations \citep{Capaccioni2015, Goesmann2015, LeRoy2015}. Molecules as large as ethanol and glycolaldehyde have also been recently detected for comet Lovejoy \citep{Biver2015}, so chemical complexity seems to be characteristic to comets. The detected abundances are fractions of a per cent relative to water, however relative to methanol the abundances are as high as few per cent. This is consistent with the modelling outcomes of this work, where complex organics are shown to be efficiently formed relative to methanol (Table~\ref{tbl:com}). This may imply that cometary complex organics formed en route in the envelope in the earliest embedded phases of star formation and were incorporated into planetesimals and cometary bodies in disc midplanes. Potentially, there are also contributions stemming from the prestellar phase, where complex organics have also been observed \citep{Requena-Torres2007, Bacmann2012, Vastel2014}. However, gas-phase processes have been proposed as the origin. Meanwhile, recent laboratory experiments indicate additional low-temperature pathways to complex organics \citep{Fedoseev2015, Chuang2016}. More detailed and quantitative comparisons between modelled midplane and observed cometary volatiles are the subject of future publications. In particular, the authors will aim to relate data on 67P obtained with the ROSINA instrument, which are free from temporal peculiarities (such as those discussed in \citealt{Qi2015}) that could potential affect remote sensing observations, with the results of these models.

\section{Conclusions}
\label{conclusions}
This paper builds on previous work by \citet{Visser2009, Visser2011} and \citet{Drozdovskaya2014} and is focused on the midplane composition of protoplanetary discs in the embedded phase. State-of-the-art physicochemical models are employed to simulate the formation of discs from the initial prestellar phase for $2.46 \times 10^{5}$~yr. Two discs are studied that vary in their respective dominant disc-growth mechanism, either viscous spreading or pure infall. Subsequently, the path of material towards the midplanes of these discs differs, predominantly in-out or simply inwards, respectively. These routes in turn set the temperatures and UV fluxes during the transport of parcels from cloud to disc. More than $100$ trajectories into each disc are computed to sample the disc midplane, and the icy content in the framework of such a dynamical model is analysed thereafter. The main conclusions are as follows:
\begin{itemize}
\item The typical main ice constituents in the midplanes are $\sim 70-80$\% H$_{2}$O, $\sim 10-20$\% CO$_{2}$, $1-10$\% NH$_{3}$ and $1-4$\% CH$_{3}$OH; however, CO$_{2}$ may dominate ($\sim 40$\%) in the outer disc when grown via pure infall.
\item Trace complex organic ices are most abundant in the outer disc ($R \gtrsim 30$~AU). Some may be as plentiful as methanol ice in the midplane, at a level of $\sim 1$\% of the total ice content, similar to that seen in comet observations. The inner disc is rich in complex organic gases.
\item The positions of snowlines of volatiles and complex organics in the midplanes of protoplanetary discs, according to relative volatilities, are retained even when dynamics and chemistry during disc formation are taken into account.
\item Dynamic infall and the chemistry en route to the midplane may enhance the amount of CO$_{2}$ and diminish CH$_{3}$OH ices in comparison to the prestellar phase. Not all volatiles are simply inherited by the midplane from the cloud. Icy planetesimals and cometary bodies reflect the provenances of the midplane ices.
\item The elevated temperatures and additional FUV photons in the envelope facilitate the formation of prebiotically-significant molecules, which may consitute as much as $\sim 10$\% of the icy mantles. Current Class II disc models may be underestimating the complex organic content, since the initial abundances set in the embedded phase could be much higher than currently assumed.
\item The C/O and C/N ratios differ between the gas and solid phases. The two ratios in the ice show little variation beyond the inner $10$~AU and both are nearly solar in the case of pure infall.
\end{itemize}
The latest theories and observations are suggesting much earlier planetesimal formation than previously thought. The results presented in this paper for the midplanes of protoplanetary discs around low-mass protostars in the embedded phase may probe the volatile and prebiotically-significant content of the pebbles that go on to feed exoplanet atmospheres and form comets. Future work will focus on deeper understanding of the consequences of these model results on cometary composition and planetary populations.

\section{Acknowledgements}
\label{acknowledgements}

This work is supported by a Huygens fellowship from Leiden University, by the European Union A-ERC grant 291141 CHEMPLAN, by the Netherlands Research School for Astronomy (NOVA) and by a Royal Netherlands Academy of Arts and Sciences (KNAW) professor prize. C.W. acknowledges support from the Netherlands Organisation for Scientific Research (NWO, program number 639.041.335). U.M. and A.T. acknowledge the National Centre for Competence in Research PlanetS supported by the Swiss National Science Foundation.

\bibliographystyle{mn2e}
\bibliography{mybib} 

\appendix

\section{Ice abundances}
\label{iceabun}

\ctable[
 star = 1,
 caption = {Mean, minimum and maximum abundances of ices beyond their respective snowlines in the spread-dominated disc at the end of the simulation ($2.46 \times 10^{5}$~yr after the onset of collapse).},
 label = tbl:iceabun_case3
 ]{@{\extracolsep{\fill}}lcllllll}{
 }{
 \hline
 \multirow{2}{*}{Species} & \multirow{2}{*}{$R_{\text{snow}}$ (AU)} & \multicolumn{3}{c}{$n \left( \text{X}_{\text{ice}} \right) / n_{\text{H}}$} & \multicolumn{3}{c}{$n \left( \text{X}_{\text{ice}} \right) / n \left( \text{H}_{2}\text{O}_{\text{ice}} \right)$} \T\\
 \cline{3-8}   &  & \multicolumn{1}{c}{mean} & \multicolumn{1}{c}{min} & \multicolumn{1}{c}{max} & \multicolumn{1}{c}{mean} & \multicolumn{1}{c}{min} & \multicolumn{1}{c}{max} \T\B\\
 \hline
 H$_{2}$O          & $~~~~~~3$ & $1.8 \left( -4 \right)$  & $1.6 \left( -5 \right)$  & $1.9 \left( -4 \right)$  & $1.0 \left( ~~~0 \right)$  & $1.0 \left( ~~~0 \right)$  & $1.0 \left( ~~0 \right)$\T\\
 CO                & $>50$     &  &  &  &  &  & \\
 CO$_{2}$          & $~~~~~~7$ & $3.6 \left( -5 \right)$  & $1.0 \left( -6 \right)$  & $3.9 \left( -5 \right)$  & $2.0 \left( -1 \right)$  & $5.9 \left( -3 \right)$  & $2.3 \left( -1 \right)$\\
 NH$_{3}$          & $~~~~~~3$ & $1.3 \left( -5 \right)$  & $1.1 \left( -5 \right)$  & $2.1 \left( -5 \right)$  & $7.2 \left( -2 \right)$  & $6.0 \left( -2 \right)$  & $1.2 \left( -1 \right)$\\
 CH$_{4}$          & $>50$     &  &  &  &  &  & \\
 CH$_{3}$OH        & $~~~~~~4$ & $5.3 \left( -6 \right)$  & $1.1 \left( -6 \right)$  & $6.2 \left( -6 \right)$  & $2.9 \left( -2 \right)$  & $6.4 \left( -3 \right)$  & $3.3 \left( -2 \right)$\\
 H$_{2}$S          & $~~~~~~7$ & $3.2 \left( -9 \right)$  & $1.1 \left( -11 \right)$ & $2.1 \left( -8 \right)$  & $1.8 \left( -5 \right)$  & $6.5 \left( -8 \right)$  & $1.2 \left( -4 \right)$\\
 N$_{2}$           & $>50$     &  &  &  &  &  & \\
 H$_{2}$CO         & $~~~~20$  & $4.9 \left( -8 \right)$  & $4.7 \left( -11 \right)$ & $1.3 \left( -7 \right)$  & $2.7 \left( -4 \right)$  & $2.6 \left( -7 \right)$  & $6.9 \left( -4 \right)$\\
 CH$_{2}$CO        & $~~~~~~9$ & $1.6 \left( -8 \right)$  & $1.9 \left( -13 \right)$ & $4.4 \left( -8 \right)$  & $9.0 \left( -5 \right)$  & $1.1 \left( -9 \right)$  & $2.4 \left( -4 \right)$\\
 HCOOH             & $~~~~~~4$ & $3.0 \left( -9 \right)$  & $1.2 \left( -19 \right)$ & $1.0 \left( -8 \right)$  & $1.6 \left( -5 \right)$  & $6.9 \left( -16 \right)$ & $5.6 \left( -5 \right)$\\
 HCOOCH$_{3}$      & $~~~~~~5$ & $1.8 \left( -7 \right)$  & $3.5 \left( -19 \right)$ & $4.0 \left( -7 \right)$  & $9.6 \left( -4 \right)$  & $2.1 \left( -15 \right)$ & $2.1 \left( -3 \right)$\\
 CH$_{3}$CHO       & $~~~~~~7$ & $4.3 \left( -10 \right)$ & $2.5 \left( -17 \right)$ & $2.6 \left( -9 \right)$  & $2.4 \left( -6 \right)$  & $1.4 \left( -13 \right)$ & $1.4 \left( -5 \right)$\\
 CH$_{3}$OCH$_{3}$ & $~~~~~~7$ & $6.7 \left( -9 \right)$  & $1.5 \left( -12 \right)$ & $2.6 \left( -8 \right)$  & $3.7 \left( -5 \right)$  & $8.5 \left( -9 \right)$  & $1.4 \left( -4 \right)$\\
 C$_{2}$H$_{5}$OH  & $~~~~~~3$ & $1.3 \left( -6 \right)$  & $1.0 \left( -8 \right)$  & $2.6 \left( -6 \right)$  & $7.2 \left( -3 \right)$  & $6.6 \left( -4 \right)$  & $1.4 \left( -2 \right)$\\
 CH$_{3}$COOH      & $~~~~~~3$ & $1.9 \left( -6 \right)$  & $9.5 \left( -8 \right)$  & $5.5 \left( -6 \right)$  & $1.5 \left( -2 \right)$  & $5.9 \left( -3 \right)$  & $4.8 \left( -1 \right)$\\
 HOCH$_{2}$CHO     & $~~~~~~3$ & $3.1 \left( -6 \right)$  & $5.6 \left( -17 \right)$ & $5.8 \left( -6 \right)$  & $1.7 \left( -2 \right)$  & $3.3 \left( -13 \right)$ & $3.2 \left( -2 \right)$\B\\
 \hline}

\ctable[
 star = 1,
 caption = {Same as Table~\ref{tbl:iceabun_case7}, but for the infall-dominated disc.},
 label = {tbl:iceabun_case7}
 ]{@{\extracolsep{\fill}}lcllllll}{
 }{
 \hline
 \multirow{2}{*}{Species} & \multirow{2}{*}{$R_{\text{snow}}$ (AU)} & \multicolumn{3}{c}{$n \left( \text{X}_{\text{ice}} \right) / n_{\text{H}}$} & \multicolumn{3}{c}{$n \left( \text{X}_{\text{ice}} \right) / n \left( \text{H}_{2}\text{O}_{\text{ice}} \right)$} \T\\
 \cline{3-8}   &  & \multicolumn{1}{c}{mean} & \multicolumn{1}{c}{min} & \multicolumn{1}{c}{max} & \multicolumn{1}{c}{mean} & \multicolumn{1}{c}{min} & \multicolumn{1}{c}{max} \T\B\\
 \hline
 H$_{2}$O          & $~~3$ & $1.2 \left( -4 \right)$  & $9.0 \left( -7 \right)$  & $1.6 \left( -4 \right)$  & $1.0 \left( ~~~0 \right)$  & $1.0 \left( ~~~0 \right)$  & $1.0 \left( ~~~0 \right)$\T\\
 CO                & $21$  & $1.8 \left( -6 \right)$  & $1.9 \left( -17 \right)$ & $1.9 \left( -5 \right)$  & $1.2 \left( -2 \right)$  & $3.0 \left( -13 \right)$ & $1.2 \left( -1 \right)$\\
 CO$_{2}$          & $~~7$ & $8.8 \left( -5 \right)$  & $2.7 \left( -6 \right)$  & $1.1 \left( -4 \right)$  & $9.7 \left( -1 \right)$  & $1.7 \left( -2 \right)$  & $3.6 \left( ~~~0 \right)$\\
 NH$_{3}$          & $~~3$ & $3.6 \left( -6 \right)$  & $5.0 \left( -9 \right)$  & $6.6 \left( -6 \right)$  & $2.7 \left( -2 \right)$  & $3.4 \left( -4 \right)$  & $4.2 \left( -2 \right)$\\
 CH$_{4}$          & $23$  & $8.4 \left( -6 \right)$  & $1.7 \left( -8 \right)$  & $1.7 \left( -5 \right)$  & $1.1 \left( -1 \right)$  & $1.7 \left( -4 \right)$  & $4.0 \left( -1 \right)$\\
 CH$_{3}$OH        & $~~3$ & $4.6 \left( -6 \right)$  & $2.2 \left( -10 \right)$ & $1.2 \left( -5 \right)$  & $3.3 \left( -2 \right)$  & $2.4 \left( -4 \right)$  & $7.8 \left( -2 \right)$\\
 H$_{2}$S          & $~~8$ & $1.0 \left( -8 \right)$  & $5.6 \left( -15 \right)$ & $2.1 \left( -8 \right)$  & $9.7 \left( -5 \right)$  & $4.7 \left( -11 \right)$ & $3.8 \left( -4 \right)$\\
 N$_{2}$           & $36$  & $6.2 \left( -6 \right)$  & $6.5 \left( -11 \right)$ & $1.5 \left( -5 \right)$  & $7.4 \left( -2 \right)$  & $1.0 \left( -6 \right)$  & $3.8 \left( -1 \right)$\\
 H$_{2}$CO         & $11$  & $3.3 \left( -7 \right)$  & $6.6 \left( -12 \right)$ & $1.5 \left( -6 \right)$  & $4.2 \left( -3 \right)$  & $5.6 \left( -8 \right)$  & $2.3 \left( -2 \right)$\\
 CH$_{2}$CO        & $~~8$ & $8.6 \left( -8 \right)$  & $9.6 \left( -10 \right)$ & $2.9 \left( -7 \right)$  & $5.9 \left( -4 \right)$  & $9.2 \left( -6 \right)$  & $1.8 \left( -3 \right)$\\
 HCOOH             & $~~3$ & $5.8 \left( -7 \right)$  & $7.9 \left( -12 \right)$ & $2.8 \left( -6 \right)$  & $5.8 \left( -3 \right)$  & $1.5 \left( -7 \right)$  & $2.8 \left( -2 \right)$\\
 HCOOCH$_{3}$      & $~~4$ & $8.0 \left( -8 \right)$  & $5.5 \left( -13 \right)$ & $3.0 \left( -7 \right)$  & $6.3 \left( -4 \right)$  & $4.7 \left( -9 \right)$  & $2.3 \left( -3 \right)$\\
 CH$_{3}$CHO       & $~~5$ & $5.2 \left( -8 \right)$  & $1.3 \left( -13 \right)$ & $6.3 \left( -17 \right)$ & $3.8 \left( -4 \right)$  & $7.8 \left( -10 \right)$ & $4.4 \left( -3 \right)$\\
 CH$_{3}$OCH$_{3}$ & $~~6$ & $5.7 \left( -8 \right)$  & $3.9 \left( -11 \right)$ & $3.6 \left( -7 \right)$  & $4.2 \left( -4 \right)$  & $1.0 \left( -6 \right)$  & $3.0 \left( -3 \right)$\\
 C$_{2}$H$_{5}$OH  & $~~3$ & $1.0 \left( -7 \right)$  & $6.7 \left( -11 \right)$ & $3.7 \left( -7 \right)$  & $7.4 \left( -4 \right)$  & $1.7 \left( -6 \right)$  & $2.3 \left( -3 \right)$\\
 CH$_{3}$COOH      & $~~3$ & $1.1 \left( -6 \right)$  & $4.8 \left( -11 \right)$ & $5.4 \left( -6 \right)$  & $1.1 \left( -2 \right)$  & $9.9 \left( -7 \right)$  & $6.9 \left( -2 \right)$\\
 HOCH$_{2}$CHO     & $~~3$ & $1 .3\left( -6 \right)$  & $3.3 \left( -12 \right)$ & $6.3 \left( -6 \right)$  & $1.1 \left( -2 \right)$  & $2.8 \left( -8 \right)$ & $2.1 \left( -1 \right)$\B\\
 \hline}


\bsp 

\label{lastpage}

\end{document}